\documentstyle[12pt]{article}

\newcommand{\mysection}{\setcounter{equation}{0}\section}
\newcommand{\Slash} {\slash \!\!\!}

\begin{document}
\vskip 0.2cm
\hfill{YITP-SB-02-32}
\vskip 0.2cm
\hfill{INLO-PUB-4/02}\\[0.5cm]
\vskip 0.2cm
\centerline{\large\bf {NLO differential distributions of massive lepton-pair 
production }}
\centerline{\large\bf {in longitudinally polarized proton-proton collisions}}
\vskip 0.4cm
\centerline {\sc V. Ravindran}
\centerline{\it Harish-Chandra Research Institute,}
\centerline{\it Chhatnag Road, Jhunsi,}
\centerline{\it Allahabad, 211019, India.}
\vskip 0.2cm
\centerline {\sc J. Smith 
\footnote{partially supported
by the National Science Foundation grant PHY-0098527.}
}
\centerline{\it C.N. Yang Institute for Theoretical Physics,}
\centerline{\it State University of New York at Stony Brook,
New York 11794-3840, USA.}
\vskip 0.2cm
\centerline {\sc W.L. van Neerven 
\footnote{Work supported
by the EC network `QCD and Particle Structure' under contract 
\hspace*{5mm} No.~FMRX--CT98--0194.}}
\centerline{\it Instituut-Lorentz}
\centerline{\it University of Leiden,}
\centerline{\it PO Box 9506, 2300 RA Leiden,}
\centerline{\it The Netherlands.}
\vskip 0.2cm
\centerline{June 2002}
\vskip 0.2cm
\centerline{\bf Abstract}
\vskip 0.3cm
We present the full next-to-leading order (NLO) corrected inclusive 
cross section 
$d^3\Delta \sigma/dQ^2/dy/dp_T$ for massive lepton pair 
production in longitudinally polarized proton-proton collisions 
$p + p\rightarrow l^+l^- + 'X'$. 
Here $'X'$ denotes any inclusive hadronic 
state and $Q$ represents the invariant mass of the lepton pair 
which has transverse momentum $p_T$ and rapidity $y$.
All QCD partonic subprocesses have been included provided the lepton pair 
is created by a virtual photon, which is a valid approximation 
for $Q<50~{\rm GeV}$.
Like in unpolarized proton-proton scattering the dominant subprocess 
is given by $q(\bar q) + g \rightarrow \gamma^* + 'X'$ so that 
massive lepton pair production provides us with an excellent 
method to measure the spin density of the gluon. 
Our calculations are carried out using the method of 
$n$-dimensional regularization by making a special choice for the
$\gamma_5$-matrix. Like in the case of many other prescriptions evanescent 
counter terms appear. They are determined by computing the NLO 
coefficient functions for $d\Delta\sigma/dQ^2$ and the polarized 
cross section for Higgs production using both $n$-dimensional 
regularization and a four dimensional regularization technique in which the 
$\gamma_5$-matrix is uniquely defined. Our calculations reveal that the 
non-singlet polarized coefficient function equals the unpolarized one up to a 
minus sign. We give predictions for double
longitudinal spin asymmetry measurements at the RHIC.

\vfill

\mysection{Introduction}
With the advent of the RHIC at BNL we have a new facility to study the 
spin structure of the proton, (for a review on the potential of RHIC 
see \cite{busa}), which supplements the existing polarized lepton-hadron
machines.  Polarized proton-proton collisions with a very high luminosity
and a maximum centre of mass energy of $\sqrt s=500~{\rm GeV}$ 
will provide us with many more details about spin distributions
than possible with the existing lepton-hadron machines, which give very 
little information about the polarized gluon and sea-quark parton densities.
An exception is photo-production of heavy flavours where the virtual
photon is almost on-shell. This process allows us to measure
the polarized gluon density and has been calculated up to
next-to-leading order (NLO) in \cite{bost1}. The gluon density can also 
be extracted from data on prompt photon production,
heavy flavour production and massive lepton pair production (Drell-Yan
or DY process) in proton-proton collisions.  The DY reaction also 
provides information on the polarized sea-quark densities.  In order to measure
these densities the above reactions have to be calculated at least up to
NLO in perturbative QCD. This has been achieved for prompt photon
production in \cite{coka}, \cite{govo} and for heavy flavour production in
\cite{bost2}. Up to the same order calculations have been done for the
polarized cross section $d^2\Delta\sigma/dQ^2/dy$ of the Drell-Yan (DY) 
process in \cite{weber} and \cite{gehr}. 
Vector boson production in NLO for the cross section
$d\Delta \sigma/dQ^2$ has been studied in \cite{kamal}. Notice that these
DY cross sections provide us with information about the sea-quark density
rather than about the gluon density. To determine the latter one needs to
study the differential distribution $d^3\Delta \sigma/dQ^2/dy/dp_T$ where the 
transverse momentum $p_T$ is sufficiently large so that the quark-gluon 
($qg$) subprocess dominates the quark-anti-quark ($q\bar q$) subprocess.
According to studies in \cite{bego1} and \cite{bego2} this already occurs 
for $p_T>Q/2$. Notice that the quark-gluon subprocess
also dominates in prompt photon production at large $p_T$. This reaction
has a much larger cross section than the one for the DY process,
which behaves like $1/Q^4$ at large $Q$ so one might favour it. However the 
former reaction also has its disadvantages, as we have already 
seen for unpolarized prompt photon production, due to
experimental and theoretical complications with photon
isolation criteria. Moreover the DY process has an additional large 
scale represented by the lepton pair mass $Q$ which turns out to be a 
much better scale to use than $p_T$. 
The NLO QCD corrections for the unpolarized DY cross section
$d^3\sigma/dQ^2/dy/dp_T$ were performed in \cite{elma} (non-singlet)
and in \cite{arre} (singlet). A complete NLO QCD calculation for the polarized
cross section does not exist yet except for the non-singlet part computed
in \cite{chco1}, \cite{chco2}. One can show, using general arguments, that
apart from a minus sign the non-singlet DY coefficient function due to the 
$q\bar q$-channel should be equal to its unpolarized analogue.

In this paper we present a complete NLO calculation of the polarized
DY cross section $d^3\Delta\sigma/dQ^2/dy/dp_T$  where all partonic 
subprocesses are included. Our findings for the non-singlet coefficient 
function agree with those obtained 
in \cite{chco1}, \cite{chco2} and they confirm the relation stated above. 
In the past
most of the corrections to physical quantities were performed using
the method of $n$-dimensional regularization \cite{hv} to define the 
singularities which occur in loop and phase space integrals. 
This method was preferred
above other regularization techniques because it manifestly preserves
the Ward-identities in non-abelian gauge field theories like QCD. 
However this advantage gets lost when one has to deal with the 
$\gamma_5$-matrix and the Levi-Civita tensor which appear in the 
electroweak standard model and in QCD when one studies polarization effects.
In this case some Ward identities are violated and have to be restored
by the addition of so-called evanescent counter terms. There exists
a large literature about the $\gamma_5$-matrix and the Levi-Civita tensor 
in the context of ultraviolet divergences (see \cite{bogi}) but a thorough
study on how to handle collinear and infrared singularities, which are 
characteristic of partonic cross sections, is still lacking.
The most often used prescription to compute polarized partonic cross sections
is called HVBM after the initials of the authors in \cite{hv} and \cite{brma}.
As is shown in \cite{larin}- \cite{neerv} this prescription also
requires that evanescent counter terms have to be taken into account 
otherwise one gets wrong coefficient functions.  
In practice it turns out that the 
computation of higher order corrections to physical quantities 
in the HVBM prescription is more complicated then the usual $n$-dimensional
regularization technique. This can be attributed to the split up of
the n-dimensional space in a $4$ and an $n-4$ dimensional subspace. 
Accordingly the gamma-matrices and the momenta have to split up which
complicates the gamma-matrix algebra and the phase space integrals.
To avoid this we introduce in this paper a prescription
for the $\gamma_5$-matrix which is easier to implement in 
computer algebra programs which do not contain a package to deal
with gamma-matrices following the HVBM prescription. Moreover the 
$n$-dimensional phase space integrals take their usual form as in
calculations where the $\gamma_5$-matrix is not present. 
The evanescent counter terms which we need
in our NLO calculation of $d^3\Delta\sigma/dQ^2/dy/dp_T$  
are extracted from an NLO computation of the DY polarized 
differential cross section $d\Delta \sigma/dQ^2$ 
and from the Higgs boson polarized total cross section.
To check our procedure we re-calculate these coefficient functions using
a four dimensional regularization technique (see e.g. \cite{hune1}) where
the $\gamma_5$-matrix is uniquely defined.

Our paper is organized as follows. In section 2 we introduce our notations
and discuss extensively the technicalities which are involved if one
wants to give a prescription for the $\gamma_5$-matrix using
$n$-dimensional regularization. We introduce our own definition and determine
the corresponding evanescent counter terms by comparing our results
with those obtained by four dimensional regularization techniques.
In section 3 we present the NLO corrections to the differential polarized
DY cross section and check the coefficient functions by re-calculating them
in a four dimensional regularization method.  In section 4
we study the NLO corrections to polarized DY production 
in proton-proton collisions at RHIC and make a 
comparison with earlier results which were
obtained in LO only. The long formulae for the soft-plus-virtual gluon
contributions to the coefficient functions can be found in Appendix A.


\pagestyle{myheadings}  
\mysection{Lowest order contributions to the polarized Drell-Yan process}
In this paper we consider the semi-inclusive Drell-Yan process
\begin{eqnarray}
\label{eqn2.1}
&& H_1(P_1,S_1)+H_2(P_2,S_2)\rightarrow \gamma^*(q) + 'X'\,,
\nonumber\\
&& \hspace*{48mm}\mid
\nonumber\\
&& \hspace*{50mm}\rightarrow l^+(l_1)+l^-(l_2)
\nonumber\\[2ex]
&& S=(P_1+P_2)^2 \,, \qquad Q^2\equiv q^2 =(l_1+l_2)^2 \,,
\end{eqnarray}
where $H_i$ ($i=1,2$) represent the incoming polarized hadrons carrying
the momenta $P_i$ and and spins $S_i$. Further $'X'$ denotes any inclusive
hadronic state which is unpolarized. The lepton pair is represented by
$l^+l^-$ with momenta $l_1$, $l_2$. In this paper we will only consider 
lepton pairs which have a sufficiently small invariant mass $Q$ so that 
the photon dominates in the above reaction and $Z$-boson exchange effects 
can be neglected. In addition to the kinematical variables
in Eq. (\ref{eqn2.1}) we need the following variables
\begin{eqnarray}
\label{eqn2.2}
T=(P_1-q)^2 \,, \qquad U=(P_2-q)^2 \,,
\end{eqnarray}
to obtain the differential cross section
\begin{eqnarray}
\label{eqn2.3}
S^2\,\frac{d^3~\Delta\sigma^{\rm H_1H_2}}{d~Q^2dTdU}(S,T,U,Q^2)
=\frac{4\pi\alpha^2}{3Q^2}\, S\,\frac{d^2~\Delta W^{\rm H_1H_2}}{dTdU}
(S,T,U,Q^2) \,.
\end{eqnarray}
In the QCD improved parton model the hadronic DY structure function
$d\Delta W^{\rm H_1H_2}$ is related to the partonic structure function
$d\Delta W_{ab}$ as follows
\begin{eqnarray}
\label{eqn2.4}
&&S\,\frac{d^2~\Delta W^{{\rm H_1H_2}}}{d~T~d~U}(S,T,U,Q^2)
\nonumber\\[2ex]
&&= \sum_{a_1,a_2=q,g}
\int_{x_{1,{\rm min}}}^1 \frac{dx_1}{x_1} \int_{x_{2,{\rm min}}}^1
\frac{dx_2}{x_2}\,\Delta f_{a_1}^{\rm H_1}(x_1,\mu^2)
\,\Delta f_{a_2}^{\rm H_2}(x_2,\mu^2)
\nonumber\\[2ex]
&&\times\,s\,\frac{d^2~\Delta W_{a_1a_2}}{d~t~d~u} (s,t,u,Q^2,\mu^2)\,.
\end{eqnarray}
In the formula above $\Delta f_a(x,\mu^2)$ ($a=q,\bar q,g$) are 
the polarized parton probability densities 
where $\mu$ denotes the 
factorization/renormalization scale and $x$ is the fraction of the 
hadron momentum carried away by the parton. The DY partonic structure 
function $d^2\Delta W_{a_1a_2}$ is computed from the partonic subprocess
\begin{eqnarray}
\label{eqn2.5}
a_1(p_1,s_1)+a_2(p_2,s_2)\rightarrow \gamma^* (q) + b_1(k_1) \cdots
b_m(k_m) 
\end{eqnarray}
and it reads
\begin{eqnarray}
\label{eqn2.6}
&& \Delta W_{a_1a_2}=K_{a_1a_2}\,\int d^4q\,\delta(q^2-Q^2)
\nonumber\\[2ex]
&&\times\prod_{i=1}^m
\int \frac{d^3k_i}{(2\pi)^3\,2E_i}\,\delta^{(4)}\left (p_1+p_2
- q-\sum_{j=1}^m k_j\right )
\nonumber\\[2ex]
&&\times |\Delta M_{a_1+a_2\rightarrow \gamma^* +b_1 \cdots b_m}|^2 \,,
\end{eqnarray}
where $K_{a_1a_2}$ denotes the colour and spin average factor and the 
polarized matrix elements are denoted $\Delta M$ (when we refer to unpolarized
structure functions, matrix elements and parton densities
we drop the $\Delta$). 
The partonic kinematical variables are defined in analogy 
with the hadronic variables in Eqs. (\ref{eqn2.1}), (\ref{eqn2.2}) as
\begin{eqnarray}
\label{eqn2.7}
s=(p_1+p_2)^2 \,, \quad t=(p_1-q)^2 
\,, \quad u=(p_2-q)^2 \,.
\end{eqnarray}
In the case parton $p_1$ emerges from hadron $H_1(P_1)$ and parton
$p_2$ emerges from hadron $H_2(P_2)$ we can establish the following relations
\begin{eqnarray}
\label{eqn2.8}
&& p_1=x_1\,P_1\,, \qquad p_2=x_2\,P_2 \,,
\nonumber\\[2ex]
&& s=x_1\,x_2\,S \,, \quad t=x_1(T-Q^2)+Q^2 \,, \quad u=x_2(U-Q^2)+Q^2\,,
\nonumber\\[2ex]
&& x_{1,{\rm min}}=\frac{-U}{S+T-Q^2}\,, \qquad
x_{2,{\rm min}}=\frac{-x_1(T-Q^2)-Q^2}{x_1S+U-Q^2}\,.
\end{eqnarray}
When parton $p_1$ emerges from hadron $H_2(P_2)$ and parton $p_2$ emerges
from $H_1(P_1)$ one obtains the same expression as in
Eq. (\ref{eqn2.4}) except that $T$ and $U$ are interchanged.
This result has to be added to Eq. (\ref{eqn2.4}). When the partonic
cross section is symmetric under $t\leftrightarrow u$ one can also use the
representation in Eq. (\ref{eqn2.4}) without adding the result where
$T$ and $U$ are interchanged provided one makes the replacement
$\Delta f_{a_1}^{\rm H_1}\Delta f_{a_2}^{\rm H_2}\rightarrow 
\Delta f_{a_1}^{\rm H_1} \Delta f_{a_2}^{\rm H_2}+ \Delta f_{a_1}^{\rm H_2}
\Delta f_{a_2}^{\rm H_1}$. Finally note
that the relation between the parton probability densities above 
and the parton momentum densities appearing
in the parton density sets in the literature or PDF libraries, which are
denoted by $\Delta f_a^{\rm PDF}(x,\mu^2)$, is given by 
$\Delta f_a^{\rm PDF}(x,\mu^2)=x~\Delta f_a(x,\mu^2)$.

When $|\Delta M_{a_1a_2}|^2$ in Eq. (\ref{eqn2.6}) is calculated up
to order $\alpha_s^2$ one encounters four partonic subprocesses
which are characterised by the two partons in their initial state.  
In the case of quarks with a mass $m\not =0$ they are given by
\begin{eqnarray}
\label{eqn2.9}
&& q(p_1,s_1)+\bar q(p_2,s_2) \rightarrow \gamma^* + 'X'\,,
\nonumber\\[2ex]
&& |\Delta M_{q\bar q}|^2=\frac{1}{4}\,{\bf Tr}\,\left (\gamma_5{\Slash}s_2\,
({\Slash}p_2-m) \,\tilde M \,\gamma_5{\Slash}s_1\,({\Slash}p_1+m)
\,\tilde M^{\dagger}\right )\,,
\\[2ex]
\label{eqn2.10}
&& q_1(\bar q_1)(p_1,s_1)+ g(p_2) \rightarrow \gamma^* + 'X'\,,
\nonumber\\[2ex]
&&|\Delta M_{qg}|^2=\frac{1}{4}\,\epsilon_{\mu\nu\lambda\sigma}\,
\frac{p_2^{\lambda} \,l_2^{\sigma}}{p_2\cdot l_2}\,{\bf Tr}\,\left 
(\tilde M^{\mu}
\,\gamma_5{\Slash}s_1\,({\Slash}p_1\pm m)\, \tilde M^{\nu\dagger}\right )\,,
\\[2ex]
\label{eqn2.11}
&& q_1(\bar q_1)(p_1,s_1)+ q_2(\bar q_2)(p_2,s_2) \rightarrow \gamma^* + 'X'
\nonumber\\[2ex]
&&|\Delta M_{q_1q_2}|^2=\frac{1}{4}\,{\bf Tr}\,\left (\gamma_5{\Slash}s_2\,
({\Slash}p_2\pm m) \,\tilde M \,\gamma_5{\Slash}s_1\,({\Slash}p_1\pm m)\,
\tilde M^{\dagger}
\right )\,,
\\[2ex]
\label{eqn2.12}
&& g(p_1)+ g(p_2) \rightarrow \gamma^* + 'X'\,,
\nonumber\\[2ex]
&&|\Delta M_{gg}|^2=
\nonumber\\[2ex]
&&\frac{1}{4}\,\epsilon_{\mu_2\nu_2\lambda_2\sigma_2}\,
\frac{p_2^{\lambda_2} \,l_2^{\sigma_2}}{p_2\cdot l_2}\,
\epsilon_{\mu_1\nu_1\lambda_1\sigma_1}\,\frac{p_1^{\lambda_1}
\,l_1^{\sigma_1}}{p_1\cdot l_1}\,{\bf Tr}\,\left (\tilde M^{\mu_1\mu_2}\,
\tilde M^{\nu_1\nu_2\dagger}\right )\,,
\end{eqnarray}
where $\tilde M$ denotes the matrix element which is given by the standard
Feynman rules. Further the symbol ${\bf Tr}$ can represent multiple
traces when the matrix elements are calculated in higher order and for
the reaction in Eq. (\ref{eqn2.11}) one can distinguish between $q_1=q_2$ and
$q_1\not =q_2$. 
The spin vectors $s_i$ and the gauge vectors $l_i$ ($i=1,2$) satisfy the 
properties
\begin{eqnarray}
\label{eqn2.13}
s_i\cdot p_i=0\,, \qquad s_i\cdot s_i=-1\,, \qquad l_i\cdot l_i=0\,.
\end{eqnarray}
When the (anti-)quark is massless then one has to make the replacements
\begin{eqnarray}
\label{eqn2.14}
\gamma_5{\Slash}s_i\,({\Slash}p_i\pm m) \rightarrow \pm\,\gamma_5\,h_i
\,{\Slash}p_i\,,
\end{eqnarray}
where $h_i$ ($i=1,2$) represent the helicities of the incoming (anti-)quarks
and the $+$ and $-$ signs on the righthand side hold for the quarks and 
anti-quarks respectively.
The definitions above are chosen in such a way that the partonic polarized
structure function satisfies the property
\begin{eqnarray}
\label{eqn2.15}
\Delta W_{a_1a_2}=W_{a_1a_2}(+,+)-W_{a_1a_2}(+,-)\,,
\end{eqnarray}
with $+,-$ denoting the helicities of the incoming partons.

The computation of the matrix elements in Eqs. (\ref{eqn2.9})-(\ref{eqn2.12})
and their virtual corrections reveals divergences which occur when the
momenta over which one integrates tend to zero (infrared), infinity 
(ultraviolet) or collinear to another momentum (collinear). The most popular
way to regularize these singularities is to choose the method of 
$n$-dimensional regularization \cite{hv} in which the space is extended to 
$n$ dimensions. The singularities are represented by pole terms of the
type $(1/\varepsilon)^k$ with $n=4+\varepsilon$. This method is very useful
because it preserves the Ward identities in the case of gauge theories.
However this is no longer the case when the $\gamma_5$ matrix 
and the Levi-Civita tensor appear like in Eqs. (\ref{eqn2.9})-(\ref{eqn2.12})
or in weak interactions. There is no consistent way to generalize
these two quantities in $n$ dimensions contrary to the ordinary 
matrix $\gamma_{\mu}$ or the metric tensor $g_{\mu\nu}$. In the
literature one has proposed various methods to extend the $\gamma_5$ matrix
and the Levi-Civita tensor to $n$ dimensions but one always needs 
so-called evanescent counter terms to restore the Ward identities.
A very popular prescription is the HVBM-scheme which was proposed in
\cite{hv} and generalised in \cite{brma}. 
In this approach the gamma-matrices and the momenta have to be split up
into a $4$ and an $n-4$ dimensional part. Therefore also the integrals
over the final state momenta have to split up in the same way. Many NLO
calculations have been done in this scheme (see e.g. \cite{bost1}-\cite{kamal},
\cite{chco1}, \cite{chco2}). However this approach requires a special
procedure to deal with the gamma-matrix algebra which is not implemented
in the program FORM \cite{form}. Since this program is used in our
calculations we prefer
another prescription for the $\gamma_5$-matrix which is given in \cite{akde}.
It gives the same results as the HVBM-scheme but it is much simpler to 
use in algebraic manipulation programs. Moreover one does not have to split 
up the integrals over the final state momenta and one can simply use the phase
space integrals computed for unpolarized reactions (see e.g. \cite{rasm}).
The procedure in \cite{akde} is given by
\begin{itemize}
\item[1.]
Replace the $\gamma_5$-matrix by
\begin{eqnarray}
\gamma_{\mu}\,\gamma_5=\frac{i}{6}\,\epsilon_{\mu\rho\sigma\tau}\,
\gamma^{\rho}\, \gamma^{\sigma}\,\gamma^{\tau}\quad \mbox {or} \quad
\gamma_5=\frac{i}{24}\,\epsilon_{\rho\sigma\tau\kappa}\,\gamma^{\rho}\,
\gamma^{\sigma}\,\gamma^{\tau}\,\gamma^{\kappa} \,.
\nonumber
\end{eqnarray}
\item[2.]
Compute all matrix elements in $n$ dimensions.
\item[3.]
Evaluate all Feynman integrals and phase space integrals in $n$-dimensions.
\item[4.]
Contract the Levi-Civita tensors in four dimensions after the Feynman 
integrals and phase space integrals are carried out.
\end{itemize}
\begin{eqnarray}
\label{eqn2.16}
\end{eqnarray}
Note that the contraction in four dimensions only applies to Lorentz
indices which are present in the Levi-Civita tensors. The last step in
(\ref{eqn2.16}) requires that one first has to apply tensorial reduction to all
integrals. For simple expressions this takes the form 
\begin{eqnarray}
\label{eqn2.17}
\int \frac{d^n~k}{(2\pi)^n}\,k^{\mu}\,k^{\nu}\,f(k,p_1,p_2)&=&A_{00}(n)\,
g^{\mu\nu} +A_{11}(n)\,p_1^{\mu}\,p_1^{\nu}+A_{22}(n)\,p_2^{\mu}\,p_2^{\nu}
\nonumber\\[2ex]
&& +A_{12}(n)\,p_1^{\mu}\,p_2^{\nu}+A_{21}(n)\,p_2^{\mu}\,p_1^{\nu}\,,
\nonumber\\[2ex]
\int dPS^{(2)} \,k^{\mu}\,k^{\nu}\,f(k,p_1,p_2)&=&B_{00}(n)\,g^{\mu\nu}
+B_{11}(n)\,p_1^{\mu}\,p_1^{\nu}+B_{22}(n)\,p_2^{\mu}\,p_2^{\nu}
\nonumber\\[2ex]
&&+B_{12}(n)\,p_1^{\mu}\,p_2^{\nu}+B_{21}(n)\,p_2^{\mu}\,p_1^{\nu}\,,
\nonumber\\[2ex]
\end{eqnarray}
where the coefficients $A_{ij},B_{ij}$ depend on $n=4 + \varepsilon$.
The HVBM-scheme or the prescription above automatically
reproduces the Adler-Bell-Jackiw anomaly (ABJ) \cite{abj} but it
violates the Ward identity for the non-singlet axial vector current
and the Adler-Bardeen theorem \cite{adba}. In order to obtain the correct 
renormalized quantities one therefore needs to invoke additional counter terms 
\cite{larin}, which are called evanescent since they do not occur for $n=4$.
This procedure was used to obtain the NLO anomalous dimensions
for the spin operators which determine the evolution of the parton spin
densities see \cite{mene}, \cite{vogel}. Notice that in our case the Drell-Yan 
process only involves one photon exchange. Therefore the $\gamma_5$-matrix
does not appear in the virtual amplitude so that there is no problem
in the ultraviolet sector.
Evanescent counter terms are also needed in cases where collinear divergences
show up like in partonic cross sections (see e.g. \cite{neerv}). This is 
characteristic for the HVBM scheme as well as any other approach similar to
the one we propose later on. The reason is that in
$n$-dimensional regularization there exists a one-to-one correspondence
between the ultraviolet divergences occurring in partonic operator matrix 
elements and the collinear divergences appearing in partonic cross sections 
provided both quantities are of twist two type. The main problem
with each prescription to extend the $\gamma_5$-matrix and the Levi-Civita
tensor beyond 4 dimensions is that one does not know beforehand which 
Ward identities are violated. Hence in principle all $\varepsilon=n-4$
terms in the matrix element might be spurious so that one is never sure to 
obtain the correct result. An example is the spurious terms occuring
in the interference terms in the quark-quark channel which are discussed
below Eq. (3.15).

A way to avoid this problem is to drop
$n$-dimensional regularization and to resort to four dimensional regularization
techniques where the $\gamma_5$-matrix anti-commutes with the other 
$\gamma_{\mu}$ matrices and the Levi-Civita tensor is clearly defined. However
four dimensional regularization methods have their drawbacks too because they
also violate Ward identities in particular in the ultraviolet sector
where it is difficult to apply techniques like Pauli-Villars \cite{pavi} 
or the introduction of an ultraviolet cut off on the divergent integrals.
In the case of collinear and infrared divergences four dimensional techniques
have been applied in a more successful way provided certain rules are
respected. In \cite{hune1} it was shown that off-shell and on-shell 
regularization leads to the same coefficient functions as those found by 
using $n$-dimensional regularization as regards the NLO unpolarized DY 
total cross section.  In \cite{maha} the same result was shown for the process 
$q+q\rightarrow q+q+\gamma^*$ where the NNLO coefficient function
agrees with the one obtained by the off-shell and on-shell techniques
applied in \cite{scne}. The off-shell and on-shell regularization techniques
are defined by (see Eqs. (\ref{eqn2.9})-(\ref{eqn2.12}))
\newpage
\begin{eqnarray}
\label{eqn2.18}
&&\mbox{{\bf off-shell}}
\nonumber\\[2ex]
&&p_i\cdot p_i=p_i^2<0,\quad m=0 \,,
\nonumber\\[2ex]
&&\mbox{$p_i$ represents the quark as well as the gluon momentum}\,,
\nonumber\\[2ex]
&&\mbox{{\bf on-shell}} 
\nonumber\\[2ex]
&&p_i\cdot p_i=m^2\not =0 \,, \quad \mbox{$p_i$ is the
quark momentum}\,,
\nonumber\\[2ex]
&& p_i\cdot p_i=0 \,, \quad \mbox{$p_i$ is the gluon momentum}\,.
\end{eqnarray}
The on-shell technique has the advantage that the partonic cross sections
are gauge invariant like the ones computed by $n$-dimensional regularization,
where the mass $m$ is set to be zero. The disadvantage is that the collinear
and infrared divergences due to the gluon have to be regularized by a different
technique. The off-shell regularization is universal both for gluons and 
quarks but the partonic cross sections depend on the chosen gauge.
However the same gauge dependence also appears in the operator matrix elements
so that it cancels in the coefficient functions after
subtraction of the former from the partonic cross sections.
This is checked for the cases mentioned above  Eq. (\ref{eqn2.18}). 
A more serious drawback is
that one has to be very cautious when collinear and infrared divergences
appear together which happens in the virtual and soft gluon 
corrections. As was shown in \cite{hune1}
the off-shell assignment as given in Eq. (\ref{eqn2.18}) leads to the
wrong coefficient function unless it is computed in the axial gauge
(see \cite{hune2}).
Therefore in the case of the virtual and soft gluon corrections one has to 
introduce a more sophisticated off-shell regularization
which respects the Kinoshita double cutting rules as formulated in \cite{kino}. 
In the case when only collinear divergences appear, which happens in
radiative processes, the off-shell regularization in Eq. (\ref{eqn2.18})
can be successfully applied.

We will use in parallel two regularization techniques to compute the 
NLO corrections to the double differential structure function in 
Eq. (\ref{eqn2.3}). 
This allows us to overcome the above problems associated with 
all the regularization techniques and with the evanescent counter terms 
which have to be introduced in the case of $n$-dimensional regularization.
For the first method 
we adopt a prescription which is slightly different from HVBM in Eq. 
(\ref{eqn2.16}), namely
\begin{itemize}
\item[1.]
Replace the $\gamma_5$-matrix in by
\begin{eqnarray}
\gamma_{\mu}\,\gamma_5=\frac{i}{6}\,\epsilon_{\mu\rho\sigma\tau}\,
\gamma^{\rho}\, \gamma^{\sigma}\,\gamma^{\tau}\quad \mbox {or} \quad
\gamma_5=\frac{i}{24}\,\epsilon_{\rho\sigma\tau\kappa}\,\gamma^{\rho}\,
\gamma^{\sigma}\,\gamma^{\tau}\,\gamma^{\kappa}\,.
\nonumber
\end{eqnarray}
\item[2.]
Compute all matrix elements in $n$ dimensions.
\item[3.]
Evaluate all Feynman integrals and phase space integrals in $n$
dimensions.
\item[4.]
Contract the Levi-Civita tensors in four dimensions after the Feynman integrals 
and soft gluon phase space integrals are done. 
\item[5.]
Contract the Levi-Civita tensors in four dimensions first before the hard
gluon phase space integrals are carried out
\end{itemize}
\begin{eqnarray}
\label{eqn2.19}
\end{eqnarray}
Note that by hard gluon phase space integrals we mean all integrals where we 
do not encounter infrared divergences.
In the subsequent part of this paper we will call this version of
$n$-dimensional regularization the RSN-scheme. It only differs from the
scheme in Eq. (\ref{eqn2.16}) regarding the procedure in contracting the
Levi-Civita tensors.
We have chosen this
approach to avoid tensorial reduction of the hard gluon phase space integrals 
in $n$ dimensions as is needed for the scheme in Eq. (\ref{eqn2.16})
which is equivalent to the HVBM-scheme.
For the virtual corrections we can simply use the method of tensorial reduction
of the Feynman integrals into scalar integrals which is presented
in \cite{pave} and \cite{been}. Notice that the soft gluon contributions
have to be treated in the same way as the virtual corrections because
they have to be added in order to cancel the final state collinear and 
infrared divergences in the final result. Because there are only a few 
soft gluon phase space integrals, their
tensorial reduction is feasible. A drawback of the RSN approach is that one 
has to compute many more evanescent counter terms than in the case of
the HVBM-scheme. These terms will be extracted from the computation
of the completely integrated NLO structure function which is one order
lower in $\alpha_s$ than the NLO correction to the double differential 
structure function in Eq. (\ref{eqn2.3}). 
The reason that $n$-dimensional integration and the contraction of the 
Levi-Civita tensors in four dimensions do not commute can be attributed to 
the metric tensor $g_{\mu\nu}$ in Eq. (\ref{eqn2.17}).  When the integration 
is performed first $g_{\mu\nu}$ acts as an $n$-dimensional object
which is indicated by the dependence of the coefficients of $A_{ij}(n)$
and $B_{ij}(n)$ in Eq. (\ref{eqn2.17}) on $n$. When the contraction  
is done first in four dimensions all metric tensors appearing in the matrix 
elements are four dimensional objects provided they have
the same Lorentz indices in common with the Levi-Civita tensors. This
explains the difference between the HVBM and the RSN scheme.
We checked that in the latter scheme $|\Delta M_{qg}|^2$ in Eq. (\ref{eqn2.10})
does not depend on the gauge vector $l_2$. This independence persists if
one adds an unpolarized gluon in the final state with gauge vector $l$ to the 
lower order processes in Eqs. (\ref{eqn2.9}), (\ref{eqn2.10})
where the sum over the physical polarizations is given by
\begin{eqnarray}
\label{eqn2.20}
&& P^{\mu\nu}(l,p)\equiv \sum_{\alpha=L,R} \epsilon^{\mu}(p,\alpha)\,
\epsilon^{\nu}(p,\alpha)=-g^{\mu\nu}+\frac{l^{\mu}\,p^{\nu}
+l^{\nu}\,p^{\mu}}{l\cdot p}\,.
\end{eqnarray}
The polarization sum above satisfies the conditions
\begin{eqnarray}
\label{eqn2.21}
 l_{\mu}\,P^{\mu\nu}= P^{\mu\nu}\,l_{\nu}=0\,,\quad l^2=0\,.
\end{eqnarray}
There is one exception. If there are two polarized gluons in the initial
state like in reaction (\ref{eqn2.12}) it turns out that $|\Delta M_{gg}|^2$
depends on $l_1$ and $l_2$ so that one has to make a specific choice for
them. In this paper we will choose $l_1=p_2$ and $l_2=p_1$.
In order to check the RSN method for the computation of the $2 \rightarrow 3$
body hard gluon processes we compute the double differential partonic 
structure function in four dimensions using the off-shell regularization 
technique in Eq. (\ref{eqn2.18}). This will be our second regularization
method.  

In the final part of this section we have to compute the evanescent
counter terms which show up in the RSN-scheme. They are extracted
from the NLO totally integrated structure function in Eq. (\ref{eqn2.6})
which has to be computed using several regularization schemes. Here we assume
that only four dimensional regularization techniques provide us with
the proper answer since the $\gamma_5$-matrix anti-commutes with the
other $\gamma_{\mu}$-matrices.
On the Born level we have the following partonic subprocesses
(see Eq. (\ref{eqn2.5}))
\begin{eqnarray}
\label{eqn2.22}
q(p_1)+\bar q(p_2) \rightarrow \gamma^* \,, \qquad 
g(p_1)+ g(p_2) \rightarrow H\,,
\end{eqnarray}
where $H$ represents a scalar particle e.g. the Higgs boson which like
the photon has a mass indicated by the same variable $Q$. The reason
that we need the latter process is that the totally integrated DY
structure function in NLO only provides
us with information about the splitting functions $\Delta P_{qq}$ and
$\Delta P_{qg}$ but we also need to know $\Delta P_{gq}$ and $\Delta P_{gg}$
which can be extracted from scalar boson production.
In lowest order the matrix elements are
\begin{eqnarray}
\label{eqn2.23}
|\Delta M_{q\bar q\rightarrow \gamma^*}^{(0)}|^2=-N\,s\,\left 
(1-\frac{\varepsilon}{2}\right )
\,, \quad |\Delta M_{gg\rightarrow H}^{(0)}|^2=-\frac{1}{8}\,(N^2-1)\,s^2\,,
\end{eqnarray}
where $N$ denotes the number of colours. The reason that the latter matrix
element has one power in $s$ more than the former can be attributed
to the effective scalar-gluon-gluon coupling which has a dimension
which is equal to the inverse of a mass. This coupling like the electromagnetic
coupling in $|\Delta M_{q\bar q\rightarrow \gamma^*}^{(0)}|^2$ has not
been included in the definition of $\Delta W$. On the Born level 
the polarized structure functions become
\begin{eqnarray}
\label{eqn2.24}
\Delta W^{(0)}_{q\bar q}=-\frac{1}{N}\,\delta(1-x)\,, 
\quad \Delta W^{(0)}_{gg}= -\frac{1}{N^2-1}\,\delta(1-x)\,,
\quad x=\frac{Q^2}{s}\,.
\end{eqnarray}
Notice that for the DY process we have divided by $1-\varepsilon/2$
whereas for scalar particle production we have removed an overall
term $Q^2/8$ in order to get unity on the right hand side of 
$\Delta W^{(0)}_{q\bar q}$ and $\Delta W^{(0)}_{gg}$ respectively.
In NLO the matrix elements in the RSN scheme as
well as in the off-shell mass scheme read
\begin{eqnarray}
\label{eqn2.25}
&&q(p_1) + \bar q(p_2) \rightarrow g(k_1) + \gamma^*(q)\,,
\nonumber\\[2ex]
|\Delta M^{(1)}_{q\bar q\rightarrow g~\gamma^*}|^2&=&-N\,C_F\,g^2\,
|\Delta T^{(1)}_{q\bar q}|^2
\nonumber\\[2ex]
|\Delta T^{(1)}_{q\bar q}|^2&=&\Bigg [\Bigg \{ \frac{4\,s\,Q^2+2\,t^2+2\,u^2}
{t\,u}-\varepsilon\,
\frac{(t+u)^2}{t\,u} \Bigg \}\Bigg (1-\frac{\varepsilon}{2}\Bigg )
\nonumber\\[2ex]
&&-4\,\varepsilon +p_1^2\,\frac{Q^2}{t^2}+p_2^2\,\frac{Q^2}{u^2}\Bigg ]\,,
\nonumber\\[2ex]
K_{q\bar q}&=&\frac{1}{N^2}\,,
\\[2ex]
\label{eqn2.26}
&&q(\bar q)(p_1) + g(p_2) \rightarrow q(\bar q)(k_1) + \gamma^*(q)\,,
\nonumber\\[2ex]
 |\Delta M^{(1)}_{qg\rightarrow q~\gamma^*}|^2&=&(N^2-1)\,T_f\,g^2\,
|\Delta T^{(1)}_{q\bar q}|^2
\nonumber\\[2ex]
|\Delta T^{(1)}_{qg}|^2&=&\Bigg [4\,\frac{Q^2}{t}-4\,\frac{Q^2}{s}
+2\,\frac{t}{s}-2\,\frac{s}{t}
\nonumber\\[2ex]
&&+\varepsilon \Bigg \{2-2\,\frac{Q^2}{t}-2\,\frac{Q^2}{s}+\frac{t}{s}
+\frac{s}{t}\Bigg \} 
+p_2^2\,\frac{2\,s\,Q^2-4\,Q^4}{s\,t^2}\Bigg ]\,,
\nonumber\\[2ex]
K_{qg}&=&\frac{1}{N\,(N^2-1)}\,,
\\[2ex]
\label{eqn2.27}
&& q(\bar q)(p_1) + g(p_2) \rightarrow q(\bar q)(k_1) + H(q)\,,
\nonumber\\[2ex]
 |\Delta M^{(1)}_{gq\rightarrow q~H}|^2&=&\frac{1}{4}\,N\,C_F\,g^2\,
\Bigg [\frac{s^2-t^2}{u}-p_1^2\,\frac{Q^4}{u^2} \Bigg ]\,,
\nonumber\\[2ex]
K_{gq}&=&\frac{1}{N\,(N^2-1)}\,,
\\[2ex]
&&g(p_1) + g(p_2) \rightarrow g(k_1) + H(q)\,,
\nonumber\\[2ex]
\label{eqn2.28}
|\Delta M^{(1)}_{gg\rightarrow g~H}|^2&=&\frac{1}{4}\,N\,(N^2-1)\,g^2\,\Bigg [
\frac{2\,s\,(s-Q^2)+4\,Q^4}{t}+\frac{2\,s\,(s-Q^2)+4\,Q^4}{u}
\nonumber\\[2ex]
&&+\frac{2\,s^3}{t\,u} +\frac{3\,(s-Q^2)^2}{s} +\frac{t^2+u^2}{s}
\nonumber\\[2ex]
&& +p_1^2\,\frac{3\,s\,Q^4-4\,Q^6}{s\,t^2}
+p_2^2\,\frac{3\,s\,Q^4-4\,Q^6}{s\,u^2}\Bigg ]\,,
\nonumber\\[2ex]
K_{gg}&=&\frac{1}{(N^2-1)^2}\,.
\end{eqnarray}
Here $g$ represents the strong coupling constant and the colour factors are 
given by
\begin{eqnarray}
\label{eqn2.29}
C_A=N\,, \qquad C_F=\frac{N^2-1}{2N}\,, \qquad T_f=\frac{1}{2}\,.
\end{eqnarray}
Notice that one has to put $p_i^2=0$ in the equations above if the matrix
elements are understood to be used in the RSN scheme where the quarks are
massless and the phase space integrals are evaluated in $n$ dimensions.
In the case of the off-shell regularization method one has to put 
$\varepsilon=0$. Also we have neglected higher powers in $p_i^2$ 
since they do not contribute to the structure function 
in the limit $p_i^2\rightarrow 0$.
Further the off-shell matrix elements are gauge dependent and we have 
taken $l^{\mu}=l^{\nu}=0$ for the polarization sum in Eq. (\ref{eqn2.19})
(Feynman gauge). In the case of the on-shell mass assignment we obtain
\begin{eqnarray}
\label{eqn2.30}
|\Delta M^{(1)}_{q\bar q\rightarrow g~\gamma^*}|^2&=&-N\,C_F\,g^2\,
\Bigg [\frac{4\,s\,Q^2+2\,t^2+2\,u^2}{(t-m^2)\,(u-m^2)}+4\,m^2\,\Bigg \{
\frac{s\,Q^2-s^2-Q^4}{s\,(t-m^2)^2}
\nonumber\\[2ex]
&&+\frac{s\,Q^2-s^2-Q^4}{s\,(u-m^2)^2}\Bigg \}\Bigg ]\,,
\\[2ex]
\label{eqn2.31}
|\Delta M^{(1)}_{qg\rightarrow q~\gamma^*}|^2&=&(N^2-1)\,T_f\,g^2
\Bigg [\frac{4\,Q^2-2\,s}{t-m^2}-\frac{4\,Q^2-2\,t}{s-m^2}
\nonumber\\[2ex]
&&-m^2\,\frac{4\,Q^2}{(t-m^2)^2}\Bigg ]\,,
\\[2ex]
\label{eqn2.32}
|\Delta M^{(1)}_{gq\rightarrow q~H}|^2&=&\frac{1}{4}\,N\,C_F\,g^2\,
\Bigg [\frac{s^2-t^2}{u-m^2} \Bigg ]\,.
\end{eqnarray}
Notice that we have omitted the on-shell matrix element 
$|\Delta M_{gg}^{{\rm on}(1)}|^2$ because the gluon is massless so there
is no regulator mass for the collinear divergences when we integrate
over the final state momenta. 

Let us first start with the computation of the structure functions 
in the off-shell and on-shell mass assignments because they immediately lead
to the correct coefficient functions in the ${\overline{\rm MS}}$-scheme.
Integration over the final state momenta $k_1$ and $q$ provides
us with the following results. For the off-shell mass assignment we get
\begin{eqnarray}
\label{eqn2.33}
\Delta \hat W_{q\bar q}^{{\rm off}(1)}&=&-a_s\,\frac{1}{N}\,C_F\,
\Bigg [\Bigg \{4\,\left ( \frac{1}{1-x} \right )_+
-2-2x\Bigg \}\Bigg \{\ln \frac{Q^2}{-p_1^2}+\ln \frac{Q^2}{-p_2^2}\Bigg \}
\nonumber\\[2ex]
&&+8\,\left (\frac{\ln (1-x)}{1-x} \right )_+-4\,(1+x)\ln(1-x)
-\frac{8(1+x^2)}{1-x}\ln x 
\nonumber\\[2ex]
&& -8+4x\Bigg ]\,,
\\[2ex]
\label{eqn2.34}
\Delta \hat W_{qg}^{{\rm off}(1)}&=&-a_s\,\frac{1}{N}\,T_f\,\Bigg [
\Bigg \{4\,x-2\Bigg \}\Bigg \{\ln \frac{Q^2}{-p_2^2} +\ln (1-x) -2\ln x\Bigg \}
\nonumber\\[2ex]
&&+(1-x)(5+3x)-2\Bigg ]\,,
\\[2ex]
\label{eqn2.35}
\Delta \hat W_{gq}^{{\rm off}(1)}&=-&a_s\,\frac{1}{N^2-1}\,C_F\,
\Bigg [\Bigg \{4-2\,x\Bigg \}\Bigg \{\ln \frac{Q^2}{-p_1^2} +\ln (1-x) 
-2\ln x\Bigg \} \nonumber\\[2ex]
&&+\frac{3}{x}(1-x)^2-2\Bigg ]\,,
\\[2ex]
\label{eqn2.36}
\Delta \hat W_{gg}^{{\rm off}(1)}&=&-a_s\,\frac{1}{N^2-1}\,C_A\,\Bigg [
\Bigg \{4\,\left (\frac{1}{1-x} \right )_+
+4-8x\Bigg \}\Bigg \{\ln \frac{Q^2}{-p_1^2}+\ln \frac{Q^2}{-p_2^2}\Bigg \}
\nonumber\\[2ex]
&&+8\,\left (\frac{\ln (1-x)}{1-x} \right )_++\Big (8-16x\Big )\ln(1-x)
-\Big (\frac{16}{1-x}+16
\nonumber\\[2ex]
&& -32x \Big )\ln x +\frac{22}{3x}\,(1-x)^3 -12+16x\Bigg ]\,,
\end{eqnarray}
where we have introduced the shorthand notation
\begin{eqnarray}
\label{eqn2.37}
a_s=\frac{\alpha_s}{4\pi}=\frac{g^2}{(4\pi)^2}\,.
\end{eqnarray}
Notice that we have omitted the soft-plus-virtual gluon corrections
in Eq. (\ref{eqn2.33}) and Eq. (\ref{eqn2.36}) which are proportional
to $\delta(1-x)$.  As was pointed out in \cite{hune1}
these terms cannot be correctly computed using the
off-shell assignment according to Eq. (\ref{eqn2.18}). 
In the case of the on-shell mass assignment we can compute the latter 
correction and we obtain
\begin{eqnarray}
\label{eqn2.38}
\Delta \hat W_{q\bar q}^{{\rm on}(1)}&=&-a_s\,\frac{1}{N}\,
C_F\,\Bigg [\Bigg \{8\,\left (
\frac{1}{1-x} \right )_+ -4-4x\Bigg \}\ln \frac{Q^2}{m^2}
\nonumber\\[2ex]
&&-\frac{4(1+x^2)}{1-x}\ln x-8\,\left (\frac{1}{1-x}\right )_+ -4+12x
\nonumber\\[2ex]
&&+\delta(1-x)\Bigg \{ 6\,\ln \frac{Q^2}{m^2} -8+8\zeta(2)\Bigg \}\Bigg ]\,,
\\[2ex]
\label{eqn2.39}
\Delta \hat W_{qg}^{{\rm on}(1)}&=&-a_s\,\frac{1}{N}\,
T_f\,\Bigg [\Bigg \{4\,x-2\Bigg \}
\Bigg \{\ln \frac{Q^2}{m^2} +2\ln (1-x) -\ln x\Bigg \}
\nonumber\\[2ex]
&&+(1-x)(5+3x)\Bigg ]\,,
\\[2ex]
\label{eqn2.40}
\Delta \hat W_{gq}^{{\rm on}(1)}&=&-a_s\,\frac{1}{N^2-1}\,
C_F\,\Bigg [\Bigg \{4-2\,x\Bigg \}
\Bigg \{\ln \frac{Q^2}{m^2} +2\ln (1-x) -3\ln x\Bigg \}
\nonumber\\[2ex]
&&+\frac{3}{x}(1-x)^2\Bigg ]\,.
\end{eqnarray}
Eqs. (\ref{eqn2.34}) and (\ref{eqn2.39}) are in agreement with \cite{mara}.
A comparison with the unpolarized structure functions $W_{q\bar q}^{(0)}$ and
$\hat W_{q\bar q}^{(1)}$ as
computed for instance in \cite{hune1} reveals the following relations
\begin{eqnarray}
\label{eqn2.41}
\Delta W_{q\bar q}^{(0)}&=&-W_{q\bar q}^{(0)} \,, \qquad 
\Delta \hat W_{q\bar q}^{{\rm off}(1)}=-\hat W_{q\bar q}^{{\rm off}(1)}\,, 
\nonumber\\[2ex]
\Delta \hat W_{q\bar q}^{{\rm on}(1)}&=&-\hat W_{q\bar q}^{{\rm on}(1)}       
+a_s\,\frac{1}{N}\,C_F\,\Big [8\,(1-x)\Big ]\,.
\end{eqnarray}
The last term is due to chiral symmetry breaking since the quark has a 
mass $m\not =0$.

We will now present the coefficient functions following from mass 
factorization denoted by
\begin{eqnarray}
\label{eqn2.42}
\Delta \hat W_{ij}=\sum_{k,l=q,g}\,\Delta \Gamma_{ki}\,\Delta \Gamma_{lj}
\Delta W_{kl}\,.
\end{eqnarray}
In the formula above all collinear divergences are absorbed by the kernels
$\Delta \Gamma_{kl}$ which depend on $\ln \mu^2/p_i^2$, 
$\ln \mu^2/m^2$
or $1/\varepsilon$ depending on the regularization method used. Here
$\mu$ represents the factorization scale which also enters the finite
DY coefficient functions $\Delta W_{kl}$. In  NLO the latter are given by
\begin{eqnarray}
\label{eqn2.43}
\Delta W_{qq}^{(1)}&=& \Delta \hat W_{qq}^{(1)}-2\,\Delta \Gamma_{qq}^{(1)}\,,
\nonumber\\[2ex]
\Delta W_{qg}^{(1)}&=& \Delta \hat W_{qg}^{(1)}-\Delta \Gamma_{qg}^{(1)}\,,
\nonumber\\[2ex]
\Delta W_{gq}^{(1)}&=& \Delta \hat W_{gq}^{(1)}-\Delta \Gamma_{gq}^{(1)}\,,
\nonumber\\[2ex]
\Delta W_{gg}^{(1)}&=& \Delta \hat W_{gg}^{(1)}-2\,\Delta \Gamma_{gg}^{(1)}\,.
\end{eqnarray}
In the case of the off- and on-shell regularization the kernels are
given by the operator matrix elements as follows
\begin{eqnarray}
\label{eqn2.44}
\Delta \Gamma_{qq}=\Delta A_{qq}\,, \quad \Delta \Gamma_{qg}=\frac{1}{2}\,
\Delta A_{qg}\,,\quad
\Delta \Gamma_{gq}=\Delta A_{gq}\,, \quad \Delta \Gamma_{gg}=\Delta A_{gg}\,.
\end{eqnarray}
For the off-shell mass assignment the latter can be found in \cite{mene}, 
\cite{masm} and they are given by
\begin{eqnarray}
\label{eqn2.45}
\Delta A_{iq}&=& \Delta A_{iq}^{{\rm PHYS}} +\Delta A_{iq}^{{\rm EOM}}\,,
\nonumber\\[2ex]
\Delta A_{ig}&=& \Delta A_{ig}^{{\rm PHYS}} \,, \qquad i=q,g\,.
\end{eqnarray}
The term $\Delta A_{iq}^{{\rm EOM}}$ is characteristic of the off-shell
mass assignment and it vanishes when the equations of motions (EOM) are
applied to the external quark indicated by the momentum $p$ so that
it does not show up in the on-shell case.
In the Feynman gauge and renormalized in the ${\overline{\rm MS}}$-scheme
they read
\begin{eqnarray}
\label{eqn2.46}
\Delta A_{qq}^{\rm PHYS,off}&=&\delta(1-x)+a_s\,C_F\,\Bigg [\Bigg \{4\,\left (
\frac{1}{1-x}\right )_+ -2-2x\Bigg \}\ln \frac{\mu^2}{-p^2}
\nonumber\\[2ex]
&&-4\,\left (\frac{\ln (1-x)}{1-x}\right )_++2\,\Big (1 + x\Big )\ln (1-x)
-\frac{2(1+x^2)}{1-x}\ln x
\nonumber\\[2ex]
&&-4 + 6x +\delta(1-x)\, \Bigg \{3\, \ln \frac{\mu^2}{-p^2}+7
-4\,\zeta(2)\Bigg \}\Bigg ]\,,
\\[2ex]
\label{eqn2.47}
\Delta A_{qq}^{\rm EOM,off}&=&a_s\,C_F\,\Bigg [-4\,x\Bigg ]\,,
\\[2ex]
\label{eqn2.48}
\Delta A_{qg}^{\rm PHYS,off}&=&a_s\,T_f\,\Bigg [\Bigg \{8x-4\Bigg \}\Bigg \{\ln
\frac{\mu^2}{-p^2} -\ln (1-x) -\ln x\Bigg \} -4 \Bigg ]\,,
\\[2ex]
\label{eqn2.49}
\Delta A_{gq}^{\rm PHYS,off}&=&a_s\,C_F\,\Bigg [\Bigg \{4-2x\Bigg \}\Bigg \{\ln
\frac{\mu^2}{-p^2} -\ln (1-x) -\ln x\Bigg \} -2 \Bigg ]\,,
\\[2ex]
\label{eqn2.50}
\Delta A_{gq}^{\rm EOM,off}&=&a_s\,C_F\,\Bigg [4\,(1-x)\Bigg ]\,,
\\[2ex]
\label{eqn2.51}
\Delta A_{gg}^{\rm PHYS,off}&=&\delta(1-x)+a_s\,C_A\,\Bigg [\Bigg \{4\,\left 
(\frac{1}{1-x} \right )_+ +4-8x\Bigg \}\ln \frac{\mu^2}{-p^2}
\nonumber\\[2ex]
&& -4\,\left (\frac{\ln (1-x)}{1-x}\right )_++\Big (-4 + 8x\Big )\ln (1-x)
-\Big (\frac{4}{1-x}
\nonumber\\[2ex]
&&+4-8x\Big )\ln x +2+\delta(1-x)
\Bigg \{\frac{11}{3}\, \ln \frac{\mu^2}{-p^2}+\frac{67}{9}
-4\,\zeta(2)\Bigg \}\Bigg ]
\nonumber\\[2ex]
&&+ a_s\,n_f\,T_f\,\delta(1-x)\Bigg [-\frac{4}{3}\,\ln \frac{\mu^2}{-p^2}
-\frac{20}{9} \Bigg ]\,.
\end{eqnarray}
The results above are calculated using the HVBM prescription according
to  Eq. (\ref{eqn2.16}). In this case the renormalization constants become
equal to
\begin{eqnarray}
\label{eqn2.52}
Z_{ij}^{\rm HVBM}=\delta_{ij}\,\delta(1-x)+a_s \left [\frac{1}{2}\,
\Delta P^{(0)}_{ij}\left (
\frac{2}{\varepsilon}+\gamma_E- \ln 4\pi \right )\right ] \quad i,j=q,g\,,
\end{eqnarray}
where $\Delta P^{(0)}_{ij}$ are the lowest order polarized splitting functions 
(see e.g. \cite{mene}, \cite{vogel}) given by
\begin{eqnarray}
\label{eqn2.53}
\Delta P^{(0)}_{qg}&=&T_f\,\left [16\,x -8 \right ]\,,
\nonumber\\[2ex]
\Delta P^{(0)}_{gq}&=&C_F\,\left [8-4\,x \right ]\,,
\nonumber\\[2ex]
\Delta P^{(0)}_{gg}&=&C_A\,\left [8\,\left (\frac{1}{(1-x)} \right )_+
+8 - 16\,x+\frac{22}{3}\,\delta(1-x) \right ]
\nonumber\\[2ex]
&& -\frac{8}{3}\,n_f\,T_f\, \delta(1-x)\,.
\end{eqnarray}
However there is one exception in the case of $Z_{qq}$. For the
HVBM prescription the latter is given by
\begin{eqnarray}
\label{eqn2.54}
Z_{qq}^{\rm HVBM}&=&\delta(1-x)+a_s \left [\frac{1}{2}\,\Delta P^{(0)}_{qq}
\left ( \frac{2}{\varepsilon}+\gamma_E- \ln 4\pi\right )\right ]
\nonumber\\[2ex]
&& -a_s\,C_F\,\Big [8\,(1-x)\Big ]\,,
\nonumber\\[2ex]
\Delta P^{(0)}_{qq}&=&C_F\,\left [8\,\left (\frac{1}{(1-x)} \right )_+
-4-4\,x+6\,\delta(1-x) \right ]\,.
\end{eqnarray}
The last term in $Z_{qq}^{\rm HVBM}$ is the evanescent counter term 
which is necessary
to ensure that the non-singlet axial vector current does not get renormalized.
This implies that the first moment of $\Delta A_{qq}^{\rm PHYS}$ must be
unity in all orders of perturbation theory. Notice that in higher order
the evanescent counter term even becomes infinite (see \cite{larin}, 
\cite{masm}). \footnote{ It appears that in the case of the naive 
$\gamma_5$-matrix prescription, where this matrix anti-commutes with the other
$\gamma$-matrices in $n$ dimensions, an evanescent counter term is not 
necessary in the non-singlet case.}  Using the same counter terms as in 
Eqs. (\ref{eqn2.52}), (\ref{eqn2.54}) one obtains the following expressions
for the operator matrix elements in the on-shell mass assignment. Using
the notation in Eq. (\ref{eqn2.45}) the operator matrix elements 
$\Delta A_{ij}^{\rm PHYS}$ become equal to 
\begin{eqnarray}
\label{eqn2.55}
\Delta A_{qq}^{\rm on}&=&\delta(1-x)+a_s\,C_F\,\Bigg [\Bigg \{4\,
\left ( \frac{1}{1-x} \right )_+ -2-2x\Bigg \}\ln \frac{\mu^2}{m^2}
\nonumber\\[2ex]
&&-8\,\left (\frac{\ln (1-x)}{1-x}\right )_+ +4\,\Big (1 + x\Big )\ln (1-x)
-4\,\left (\frac{1}{1-x}\right )_+
\nonumber\\[2ex]
&&-2 + 6x +\delta(1-x)\, \Bigg \{3\, \ln \frac{\mu^2}{m^2}+4 \Bigg \}\Bigg ]\,,
\\[2ex]
\label{eqn2.56}
\Delta A_{qg}^{\rm on}&=&a_s\,T_f\,\Bigg [\Bigg \{8x -4\Bigg \}\,
\ln \frac{\mu^2}{m^2} \Bigg ]\,,
\\[2ex]
\label{eqn2.57}
\Delta A_{gq}^{\rm on}&=&a_s\,C_F\,\Bigg [\Bigg \{4-2x\Bigg \}\Bigg \{\ln
\frac{\mu^2}{m^2} -2\ln x\Bigg \} +4\,(1 -x)\Bigg ]\,,
\end{eqnarray}
and $\Delta A_{iq}^{\rm EOM}=0$.
Notice that the first moment of $\Delta A_{qq}^{\rm on}$ in Eq. (\ref{eqn2.55})
is still not equal to unity
even after subtraction of the evanescent counter term. This is because
the non-singlet axial vector current is not conserved for massive quarks 
and therefore has to undergo a finite renormalization. This additional term,
which is equal to $a_s\,C_F\,4\,(1-x)$, will compensate the last term in Eq.
(\ref{eqn2.41}) while computing the difference in Eq. (\ref{eqn2.43}). 
After mass factorization one obtains the same coefficient functions
irrespective whether the off-shell or the on-shell regularization is used.
The results are
\begin{eqnarray}
\label{eqn2.58}
\Delta W_{q\bar q}^{(1)}&=&-a_s\,\frac{1}{N}\,C_F\,\Bigg [\Bigg \{8\,\left (
\frac{1}{1-x} \right )_+ -4-4x\Bigg \} \ln \frac{Q^2}{\mu^2}
\nonumber\\[2ex]
&&+16\,\left (\frac{\ln (1-x)}{1-x}
\right )_+-8\,(1+x)\,\ln (1-x) -\frac{4(1+x^2)}{1-x}\ln x
\nonumber\\[2ex]
&&+\delta(1-x)\Bigg \{
6\,\ln \frac{Q^2}{\mu^2} -16+8\,\zeta(2)\Bigg \}\Bigg ]\,,
\\[2ex]
\label{eqn2.59}
\Delta W_{qg}^{(1)}&=&-a_s\,\frac{1}{N}\,T_f\,\Bigg [\Bigg \{4\,x-2\Bigg \}
\Bigg \{ \ln\frac{Q^2}{\mu^2} +2\,\ln (1-x) -\ln x\Bigg \}
\nonumber\\[2ex]
&& +(1-x)(5+3x)\Bigg ]\,,
\\[2ex]
\label{eqn2.60}
\Delta W_{gq}^{(1)}&=&-a_s\,\frac{1}{N^2-1}\,C_F\,\Bigg [\Bigg \{4-2\,x\Bigg \}
\Bigg \{ \ln \frac{Q^2}{\mu^2} +2\ln (1-x) -\ln x \Bigg \}
\nonumber\\[2ex]
&&+\frac{3}{x}(1-x)^2 -4\,(1-x)\Bigg ]\,,
\\[2ex]
\label{eqn2.61}
\Delta W_{gg}^{(1)}&=&-a_s\,\frac{1}{N^2-1}\,C_A\,
\Bigg [\Bigg \{8\,\left (
\frac{1}{1-x} \right )_+ +8-16x\Bigg \} \ln \frac{Q^2}{\mu^2}
\nonumber\\[2ex]
&& +16\,\left (\frac{\ln (1-x)}{1-x} \right )_++\Big (16-32x\Big )\ln(1-x)
\nonumber\\[2ex]
&& -\Big (\frac{8}{1-x}+8-16x \Big )\ln x +\frac{22}{3x}\,(1-x)^3 -16\,(1-x)
\nonumber\\[2ex]
&& +\delta(1-x) \Bigg \{\frac{22}{3} + 8\,\zeta(2) \Bigg \} \Bigg ]\,.
\end{eqnarray}
Notice that the coefficient of the $\delta(1-x)$ term in Eq. (\ref{eqn2.58})
can only be obtained from the on-shell scheme. The latter scheme could
not be applied to the process $g+g \rightarrow g + H$ in Eq. (\ref{eqn2.28})
so that the $\delta(1-x)$-term in Eq. (\ref{eqn2.61}) can only be inferred
from $n$-dimensional regularization as HVBM or RSN.
Furthermore from Eq. (\ref{eqn2.58}) one derives
\begin{eqnarray}
\label{eqn2.62}
\Delta W_{q\bar q}^{(1)}&=&-W_{q\bar q}^{(1)}\,.
\end{eqnarray}
This relation, which only holds for coefficient functions, has to be satisfied
irrespective which regularization scheme is chosen.

Now we want to compute the coefficients above using the RSN approach presented
in Eq. (\ref{eqn2.19}) and determine the corresponding evanescent counter
terms. The DY partonic structure functions are given by
\begin{eqnarray}
\label{eqn2.63}
\Delta \hat W_{q\bar q}^{{\rm RSN},(1)}&=&\Delta W_{q\bar q}^{(1)}
-a_s\,\frac{1}{N}\,\left [\Delta P^{(0)}_{qq}\left (
\frac{2}{\varepsilon}+\gamma_E- \ln 4\pi \right )\right ]
\nonumber\\[2ex]
&&+ a_s\,\frac{1}{N}\,C_F\, \Big [8\,(1-x)\Big ] \,,
\\[2ex]
\label{eqn2.64}
\Delta \hat W_{qg}^{{\rm RSN},(1)}&=&\Delta W_{qg}^{(1)}-a_s\,\frac{1}{N}\,
\left [\frac{1}{4}\,\Delta P^{(0)}_{qg}\left (
\frac{2}{\varepsilon}+\gamma_E- \ln 4\pi \right )\right ]
\nonumber\\[2ex]
&& + a_s\,\frac{1}{N}\,T_f\,\Big [4\,(1-x) \Big ]\,,
\\[2ex]
\label{eqn2.65}
\Delta \hat W_{gq}^{{\rm RSN},(1)}&=&\Delta W_{gq}^{(1)} -a_s\,\frac{1}{N^2-1}\,
\left [\frac{1}{2}\,\Delta P^{(0)}_{gq}\left ( \frac{2}{\varepsilon}+\gamma_E
- \ln 4\pi \right ) \right ]
\nonumber\\[2ex]
&&  - a_s\,\frac{1}{N^2-1}\,C_F\,\Big [4\,(1-x) \Big ]\,,
\\[2ex]
\label{eqn2.66}
\Delta \hat W_{gg}^{{\rm RSN},(1)}&=&\Delta W_{gg}^{(1)} 
-a_s\,\frac{1}{N^2-1}\,\left [\Delta P^{(0)}_{gg}\left (
\frac{2}{\varepsilon}+\gamma_E- \ln 4\pi \right ) \right ]
\nonumber\\[2ex]
&& -a_s\,\frac{1}{N^2-1}\,C_A\,\Big [16\,(1-x) \Big ]\,.
\end{eqnarray}
This scheme violates the relation in Eq. (\ref{eqn2.62}) in a similar
way as the on-shell result in Eq. (\ref{eqn2.41}). This is because
the $n$-dimensional extension of the $\gamma_5$-matrix breaks the chiral 
symmetry. However this is unphysical because the quarks are still massless.
The kernels for the RSN prescription follow from the 
requirement that the RSN-scheme must lead to the same coefficient functions
as the four dimensional regularization methods. From mass factorization (see
Eq. (\ref{eqn2.43})) and the coefficient functions presented in Eqs.
(\ref{eqn2.58})-(\ref{eqn2.61}) we infer the kernels
\begin{eqnarray}
\label{eqn2.67}
\Delta \Gamma_{qq}^{\rm RSN}&=&\delta(1-x)+a_s \left [\frac{1}{2}\,
\Delta P^{(0)}_{qq} \left ( \frac{2}{\varepsilon}+\gamma_E
- \ln 4\pi\right )\right ]
\nonumber\\[2ex]
&& -a_s\,C_F\,\Big [4\,(1-x)\Big ]\,,
\\[2ex]
\label{eqn2.68}
\Delta \Gamma_{qg}^{\rm RSN}&=&a_s \left [\frac{1}{4}\,\Delta P^{(0)}_{qg}
\left ( \frac{2}{\varepsilon}+\gamma_E- \ln 4\pi\right )\right ]
\nonumber\\[2ex]
&& -a_s\,T_f\,\Big [4\,(1-x)\Big ]\,,
\\[2ex]
\label{eqn2.69}
\Delta \Gamma_{gq}^{\rm RSN}&=&a_s \left [\frac{1}{2}\,\Delta P^{(0)}_{gq}
\left ( \frac{2}{\varepsilon}+\gamma_E- \ln 4\pi\right )\right ]
\nonumber\\[2ex]
&& +a_s\,C_F\,\Big [4\,(1-x)\Big ]\,,
\\[2ex]
\label{eqn2.70}
\Delta \Gamma_{gg}^{\rm RSN}&=&\delta(1-x)+a_s \left [
\frac{1}{2}\,\Delta P^{(0)}_{gg} \left ( \frac{2}{\varepsilon}+\gamma_E
- \ln 4\pi\right )\right ]
\nonumber\\[2ex]
&& +a_s\,C_A\,\Big [8\,(1-x)\Big ]\,,
\end{eqnarray}
where the additional terms are characteristic of the RSN-scheme.

Finally we want to show that the prescription given in Eq. (\ref{eqn2.16})
leads to the same results as the conventional HVBM approach in \cite{hv}
and \cite{brma}. A straightforward calculation gives
\begin{eqnarray}
\label{eqn2.71}
\Delta \hat W_{q\bar q}^{{\rm HVBM},(1)}&=&\Delta W_{q\bar q}^{(1)}
-a_s\,\frac{1}{N}\,\left [\Delta P^{(0)}_{qq}\left (
\frac{2}{\varepsilon}+\gamma_E- \ln 4\pi \right )\right ]
\nonumber\\[2ex]
&&+ a_s\,\frac{1}{N}\,C_F\, \Big [16\,(1-x)\Big ]\,,
\\[2ex]
\label{eqn2.72}
\Delta \hat W_{qg}^{{\rm HVBM},(1)}&=&\Delta W_{qg}^{(1)}
-a_s\,\frac{1}{N}\,\left [\frac{1}{4}\,\Delta P^{(0)}_{qg}\left (
\frac{2}{\varepsilon}+\gamma_E- \ln 4\pi \right )\right ]\,,
\\[2ex]
\label{eqn2.73}
\Delta \hat W_{gq}^{{\rm HVBM},(1)}&=&\Delta W_{gq}^{(1)}
-a_s\,\frac{1}{N^2-1}\,\left [\frac{1}{2}\,\Delta P^{(0)}_{qq}\left (
\frac{2}{\varepsilon}+\gamma_E- \ln 4\pi \right )\right ]\,,
\nonumber\\
\\[2ex]
\label{eqn2.74}
\Delta \hat W_{gg}^{{\rm HVBM},(1)}&=&\Delta W_{q\bar q}^{(1)}
-a_s\,\frac{1}{N^2-1}\,\left [\Delta P^{(0)}_{gg}\left (
\frac{2}{\varepsilon}+\gamma_E- \ln 4\pi \right )\right ]\,.
\end{eqnarray}
The results in Eqs. (\ref{eqn2.71}) and (\ref{eqn2.72}) agree with those
in \cite{weber} which were obtained by the conventional HVBM approach.
The kernels for the HVBM prescription are obtained according to the same
requirement as used for the RSN-approach above Eq. (\ref{eqn2.67}) and they
read
\begin{eqnarray}
\label{eqn2.75}
\Delta \Gamma_{qq}^{\rm HVBM}&=&\delta(1-x)+a_s \left [
\frac{1}{2}\,\Delta P^{(0)}_{qq}\left ( \frac{2}{\varepsilon}+\gamma_E
- \ln 4\pi\right )\right ]
\nonumber\\[2ex]
&& -a_s\,C_F\,\Big [8\,(1-x)\Big ]\,,
\\[2ex]
\label{eqn2.76}
\Delta \Gamma_{qg}^{\rm HVBM}&=&a_s \left [
\frac{1}{4}\,\Delta P^{(0)}_{qg}\left ( \frac{2}{\varepsilon}+\gamma_E
- \ln 4\pi\right )\right ]\,, 
\\[2ex]
\label{eqn2.77}
\Delta \Gamma_{gq}^{\rm HVBM}&=&a_s \left [
\frac{1}{2}\,\Delta P^{(0)}_{gq}\left ( \frac{2}{\varepsilon}+\gamma_E
- \ln 4\pi\right )\right ]\,,
\\[2ex]
\label{eqn2.78}
\Delta \Gamma_{gg}^{\rm HVBM}&=&\delta(1-x)+a_s \left [
\frac{1}{2}\,\Delta P^{(0)}_{gg}\left ( \frac{2}{\varepsilon}+\gamma_E
- \ln 4\pi\right )\right ]\,. 
\end{eqnarray}
As in the case of unpolarized quantities the kernels above are equal to the 
operator renormalization constants in Eqs. (\ref{eqn2.52}), (\ref{eqn2.53}),
except that $\Delta \Gamma_{qg}^{\rm HVBM}=1/2~\Delta Z_{qg}^{\rm HVBM}$
(see also Eq. (\ref{eqn2.44})),
provided the same $\gamma_5$-matrix prescription is used for ultraviolet
and collinear divergences. Further notice that RSN and HVBM schemes
only affect the regular part of the kernels and do not alter 
the singular part represented by $\delta(1-x)$.

Summarizing the results obtained in this section we have introduced a
new scheme, called RSN, which is designed to regularize the 
hard gluon contributions to the double differential structure functions
in Eq. (\ref{eqn2.4}) when the latter are computed in NLO. For that purpose 
we calculated the
evanescent counter terms by the requirement that in the 
${\overline {\rm MS}}$ scheme one should obtain the same coefficient functions
as obtained by four dimensional regularization methods. One of the most
important results is the relation between the polarized and unpolarized
structure functions in Eq. (\ref{eqn2.62}) which has to be satisfied
irrespective of the prescription used for the $\gamma_5$-matrix.
That this relation has to emerge from the calculations can be simply inferred
from the off-shell regularization technique (see the result for $W_{q\bar q}$
in Eq. (\ref{eqn2.41})). In four dimensions the two $\gamma_5$-matrices 
in the non-singlet part of the matrix element in Eq. (\ref{eqn2.9})
can be brought together since they are present in the same trace so that
they yield unity. In this way the matrix element becomes the same
as for unpolarized reactions. 

\mysection{NLO corrections to polarized lepton-pair production}
After having specified the RSN-prescription for the $\gamma_5$ matrix and
the Levi-Civita tensor in the previous section we present the NLO
corrections to the quantity 
\begin{eqnarray}
\label{eqn3.1}
s\,\frac{d^2~{\Delta {\hat W}^{(1)}}_{a_1a_2}}{d~t~d~u}&=&
K_{a_1a_2}\,\frac{S_{\varepsilon}}{(4\pi)^2\,\Gamma(1+\varepsilon/2)}\,
\left (\frac{t~u}{Q^2~s}\right )^{\varepsilon/2}
\,\delta(s+t+u-Q^2)
\nonumber\\[2ex]
&&\times |\Delta M_{a_1a_2\rightarrow b_1~\gamma^*}^{(1)}|^2\,,
\quad \mbox{with}\quad n=4+\varepsilon\,.
\end{eqnarray}
Here $\Gamma(x)$ denotes the gamma-function and the spherical factor 
$S_{\varepsilon}$ is defined by
\begin{eqnarray}
\label{eqn3.2}
S_{\varepsilon}=\exp\left (\frac{\varepsilon}{2}\Big (\gamma_E
-\ln 4\pi\Big )\right )\,.
\end{eqnarray}
Notice that the expression in Eq. (\ref{eqn3.1}) holds for the Born
reactions in Eq. (\ref{eqn2.25}), (\ref{eqn2.26}) as well as for the 
virtual corrections presented below. The calculation proceeds in the same way 
as e.g. done in the case of unpolarized
processes like heavy flavour production \cite{beku} or Higgs production
\cite{rasm}. Starting with the virtual contributions we compute the one-loop
corrections to the reactions in Eqs. (\ref{eqn2.25}), (\ref{eqn2.26}).
The Feynman graphs can be found in Fig 5 of \cite{maha}.
First we perform tensorial reduction of the Feynman integrals 
into scalar integrals following
the procedure in \cite{pave}, \cite {been}. The scalar integrals
can be found in Appendix D of \cite{mama} and we use $n$-dimensional
regularization for the ultraviolet, infrared and collinear divergences. 
Finally we contract the Levi-Civita tensors in four dimensions.
The results which hold in the HVBM- as well as RSN-schemes are given by
\begin{eqnarray}
\label{eqn3.3}
s\,\frac{d^2~{\Delta {\hat W}}_{q\bar q}^{\rm V}}
{d~t~d~u}&=& \delta(s+t+u-Q^2)\,
S_{\varepsilon}^2\,a_s^2\, \frac{1}{N}
\nonumber\\[2ex]
&&\times \Bigg [
C_F(C_F-C_A/2)\,\Bigg \{\frac{8}{\varepsilon^2}+\left (8\,\ln \frac{-t}{\mu^2}
-6\right )\frac{1}{\varepsilon}
\nonumber\\[2ex]
&& -4\,{\rm Li}_2\left (\frac{s-Q^2}{s} \right )-4\,{\rm Li}_2 \left (
\frac{t}{Q^2}\right ) +2\,\ln \frac{-t}{\mu^2}\ln \frac{-u}{\mu^2}
\nonumber\\[2ex]
&& +4\,\ln \frac{-t}{\mu^2} \ln \frac{s}{\mu^2} -2\,\ln^2\frac{s}{\mu^2}
+2\,\ln^2\left (\frac{t-Q^2}{t} \right )
\nonumber\\[2ex]
&&-2\,\ln^2\left( \frac{Q^2-t}{Q^2}\right )-6\,\ln \frac{-t}{\mu^2}
+3\,\ln \frac{s}{Q^2} -8\,\zeta(2)+8 
\nonumber\\[2ex]
&&+ \Big (t \Leftrightarrow u\Big ) \Bigg \}
\times |\Delta T^{(1)}_{q\bar q}|^2
\nonumber\\[2ex]
&&+C_AC_F\,\Bigg \{\frac{8}{\varepsilon^2}+\left (12\,\ln \frac{-t}{\mu^2}
-4\,\ln \frac{s}{\mu^2}-3\right )\frac{1}{\varepsilon}
-2\,{\rm Li}_2 \left (\frac{t}{Q^2}\right )
\nonumber\\[2ex]
&&+5\,\ln \frac{-t}{\mu^2}\ln \frac{-u}{\mu^2}-6\ln \frac{-t}{\mu^2}
\ln \frac{s}{\mu^2} +4\,\ln^2 \frac{-t}{\mu^2}+\ln^2 \frac{s}{\mu^2}
\nonumber\\[2ex]
&&+\ln^2\left (\frac{t-Q^2}{t}\right )-\ln^2\left( \frac{Q^2-t}{Q^2}\right )
-3\ln \frac{-t}{\mu^2}+\frac{3}{2}\,\ln \frac{s}{Q^2}
\nonumber\\[2ex]
&&-7\,\zeta(2)+4 +\Big (t \Leftrightarrow u\Big )  \Bigg \}
|\Delta T^{(1)}_{q\bar q}|^2
\nonumber\\[2ex]
&&+C_F(C_F-C_A/2)\,\Bigg \{\frac{16s+8t}{u}\,\Bigg ({\rm Li}_2 \left
(\frac{t}{Q^2} \right ) + {\rm Li}_2\left (\frac{s-Q^2}{s} \right )
\nonumber\\[2ex]
&&+\frac{1}{2}\,\ln^2\frac{-t}{s}-\frac{1}{2}\,\ln^2\left (\frac{t-Q^2}{t}
\right )+\frac{1}{2}\,\ln^2\left( \frac{Q^2-t}{Q^2}\right ) \Bigg )
\nonumber\\[2ex]
&&+\frac{8s(s-2Q^2)}{(s-Q^2)^2}\,\ln\frac{s}{Q^2}
+\Bigg (\frac{4tu}{(Q^2-t)^2}
-\frac{8(t-2u)}{Q^2-t}-16\Bigg )
\nonumber\\[2ex]
&&\times  \ln \frac{-t}{Q^2}+\frac{4(s+u)}{t}+\frac{8s}{s-Q^2}
+\frac{4u}{Q^2-t}-4 +\Big (t \Leftrightarrow u\Big )\Bigg \}
\nonumber\\[2ex]
&& +C_AC_F\,\Bigg \{\Bigg ( \frac{2tu}{(Q^2-t)^2}+\frac{8(u-t)}{Q^2-t}-8\Bigg )
\ln \frac{-t}{Q^2}-\frac{2(s+u)}{t}
\nonumber\\[2ex]
&&+\frac{2u}{Q^2-t}-2 +\Big (t \Leftrightarrow u\Big )\Bigg \}
 \Bigg ]\,.
\end{eqnarray}
Apart from an overall minus sign the above result agrees with the unpolarized
virtual contribution in \cite{elma}. Note that $\varepsilon$ terms in
$|\Delta T^{(1)}_{q\bar q}|^2$ (see Eq. (\ref{eqn2.25})) differ from
those in the unpolarized case, denoted by $|T^{(1)}_{q\bar q}|^2$, which 
causes the breakdown of the relation in Eq. (\ref{eqn2.62}). Also we find
\begin{eqnarray}
\label{eqn3.4}
s\,\frac{d^2~{\Delta {\hat W}}^{\rm V}_{qg}}
{d~t~d~u}&=& \delta(s+t+u-Q^2)\,
S_{\varepsilon}^2\,a_s^2\,
\frac{1}{N}
\nonumber\\[2ex]
&&\times \Bigg [C_F\,\Bigg \{-\frac{8}{\varepsilon^2}+\left (-4\,
\ln \frac{-t}{\mu^2}-8\,\ln \frac{-u}{\mu^2}+4\,\ln \frac{s}{\mu^2} + 6
\right ) \frac{1}{\varepsilon}
\nonumber\\[2ex]
&&+2\,{\rm Li}_2 \left (\frac{t}{Q^2} \right )+2\,{\rm Li}_2 \left (
\frac{u}{Q^2} \right ) -6\,\ln \frac{-t}{\mu^2}\ln \frac{-u}{\mu^2}
\nonumber\\[2ex]
&&+2\,\ln \frac{-t}{\mu^2} \ln \frac{s}{\mu^2}
+4\,\ln \frac{-u}{\mu^2}\ln \frac{s}{\mu^2}
-3\,\ln^2 \frac{-u}{\mu^2}-\ln^2 \frac{s}{\mu^2}
\nonumber\\[2ex]
&&-\ln^2\left (\frac{t-Q^2}{t}\right )
-\ln^2\left (\frac{u-Q^2}{u}\right )
+\ln^2\left( \frac{Q^2-t}{Q^2}\right )
\nonumber\\[2ex]
&&+\ln^2\left( \frac{Q^2-u}{Q^2}\right )
+3\,\ln \frac{tu}{s\mu^2}+3\,\ln \frac{Q^2}{\mu^2}+12\,\zeta(2)-9 \Bigg \}
\nonumber\\[2ex]
&&\times |\Delta T^{(1)}_{qg}|^2
\nonumber\\[2ex]
&&+C_A\,\Bigg \{ -\frac{4}{\varepsilon^2}-\left (4\,\ln \frac{-t}{\mu^2}
\right )\frac{1}{\varepsilon}
+{\rm Li}_2\left (\frac{s-Q^2}{s} \right )
-{\rm Li}_2 \left (\frac{u}{Q^2}\right )
\nonumber\\[2ex]
&&-2\,\ln^2\frac{-t}{\mu^2}+\frac{1}{2}\,\ln^2 \frac{u}{s}
+\frac{1}{2}\,\ln^2\left (\frac{u-Q^2}{u}\right )
\nonumber\\[2ex]
&&-\frac{1}{2}\,\ln^2\left( \frac{Q^2-u}{Q^2}\right )
-\zeta(2)+1 \Bigg \} |\Delta T^{(1)}_{qg}|^2
\nonumber\\[2ex]
&&+C_F\,\Bigg \{ \frac{4(t+2u)}{s}\,\Bigg (-{\rm Li}_2
\left (\frac{s-Q^2}{s} \right )+{\rm Li}_2 \left (\frac{t}{Q^2}\right )
\nonumber\\[2ex]
&&-\ln\frac{-u}{\mu^2} \ln\frac{-t}{s}+\frac{1}{2}\,\ln^2 \frac{-t}{\mu^2}
-\frac{1}{2}\,\ln^2 \frac{s}{\mu^2}-\frac{1}{2}\,\ln^2\left
(\frac{t-Q^2}{t}\right )
\nonumber\\[2ex]
&& +\frac{1}{2}\,\ln^2\left( \frac{Q^2-t}{Q^2}\right )+5\,\zeta(2)\Bigg )
+\Bigg (\frac{2su}{(s-Q^2)^2} -\frac{2(s+4u)}{s-Q^2}\Bigg )
\nonumber\\[2ex]
&& \times \ln \frac{s}{Q^2}
+\Bigg (\frac{2st}{(Q^2-t)^2}+\frac{4(2s-t)}{Q^2-t}-8\Bigg )\ln \frac{-t}{Q^2}
-\frac{2u}{s}+\frac{2u}{t}
\nonumber\\[2ex]
&&+\frac{2s}{Q^2-t}-\frac{2u}{s-Q^2}-2 \Bigg \}
\nonumber\\[2ex]
&&+C_A\,\Bigg \{ \frac{2(t+2u)}{s}\,\Bigg ({\rm Li}_2\left (\frac{s-Q^2}{s}
\right )-{\rm Li}_2 \left (\frac{t}{Q^2}\right )
\nonumber\\[2ex]
&&+\ln\frac{-u}{\mu^2} \ln\frac{-t}{s}-\frac{1}{2}\,\ln^2 \frac{-t}{\mu^2}
+\frac{1}{2}\,\ln^2 \frac{s}{\mu^2}+\frac{1}{2}\,\ln^2\left
(\frac{t-Q^2}{t}\right )
\nonumber\\[2ex]
&&-\frac{1}{2}\,\ln^2\left( \frac{Q^2-t}{Q^2}\right )-5\,\zeta(2)\Bigg )
-\frac{2s}{s-Q^2} \ln \frac{s}{Q^2}
\nonumber\\[2ex]
&& -\frac{2t}{Q^2-t}\ln \frac{-t}{Q^2}+\frac{2u}{s}-\frac{2u}{t}\Bigg \}
\Bigg ]\,.
\end{eqnarray}
Here the function ${\rm Li}_2(x)$ in the expressions above denotes the 
di-logarithm which e.g. can be found in \cite{lewin}. Next we compute
all $2 \rightarrow 3$-body processes which contribute to the structure
function
\begin{eqnarray}
\label{eqn3.5}
s\,\frac{d^2~{\Delta {\hat W}}^{(2)}_{a_1a_2}}
{d~t~d~u}&=&\frac{1}{2\pi} K_{a_1a_2}\,\frac{S_{\varepsilon}^2}
{(4\pi)^4\Gamma(1+\varepsilon)}\,
\left (\frac{t~u-Q^2~s_4}{\mu^2~s}\right )^{\varepsilon/2}\,\left (
\frac{s_4}{\mu^2}\right)^{\varepsilon/2}
\nonumber\\[2ex]
&&\times \overline{|\Delta M^{(2)}_{a_1a_2\rightarrow b_1b_2~\gamma^*}|^2}\,,
\nonumber\\[2ex]
&& \mbox{with} \quad s_4=s+t+u-Q^2\,,
\end{eqnarray}
where $\overline{|\Delta M^{(2)}_{a_1a_2\rightarrow b_1b_2~\gamma^*}|^2}$ is 
the second order
matrix element integrated over the polar angle $\theta_1$ and the
azimuthal angle $\theta_2$ so 
\begin{eqnarray}
\label{eqn3.6}
&& \overline{|\Delta M^{(2)}_{a_1a_2\rightarrow b_1b_2~\gamma^*}|^2}
\nonumber\\[2ex]
&&=\int_0^{\pi} d\theta_1\, (\sin \theta_1)^{1+\varepsilon}\int_0^{\pi}
d\theta_2\, (\sin \theta_2)^{\varepsilon}\,
|\Delta M^{(2)}_{a_1a_2\rightarrow b_1b_2~\gamma^*}(\theta_1,\theta_2)|^2
\nonumber\\[2ex]
&&\equiv \int d\Omega_{n-1}
|\Delta M^{(2)}_{a_1a_2\rightarrow b_1b_2~\gamma^*}(\theta_1,\theta_2)|^2\,.
\end{eqnarray}
The Feynman graphs for the $2 \rightarrow 3$ body parton subprocesses
can be found in Figs. 6-9 in \cite{maha}. These reactions have been calculated
using two different regularization methods for the collinear divergences.
In $n$ dimensions we use the RSN prescription mentioned in Eq. (\ref{eqn2.19}).
In this case the collinear divergences in Eq. (\ref{eqn3.6}) manifest 
themselves as pole terms
of the type $1/\varepsilon$. For $n=4$ we adopted the off shell regularization
method in Eq. (\ref{eqn2.18}) where the collinear divergences show up as 
logarithms $\ln s/p^2$. Notice that the off-shell regularization technique
is gauge dependent and we have computed Eq. (\ref{eqn3.6}) in the
Feynman gauge (see the comments below Eq. (\ref{eqn2.29})). The gauge
dependent terms will be cancelled after mass factorization via 
the operator matrix elements in Eqs. (\ref{eqn2.45})-(\ref{eqn2.50}) which
are computed in the same gauge.
Decomposing the matrix element
in Eq. (\ref{eqn3.6}) into colour factors one obtains the following results
for the non-singlet $q\bar q$ and the singlet $qg$ subprocesses 
\begin{eqnarray}
\label{eqn3.7}
&&q + \bar q \rightarrow g + g + \gamma^*\,,
\nonumber\\[2ex]
\overline{|\Delta M^{(2)}_{q\bar q\rightarrow gg~\gamma^*}|^2}&=&
g^4\,e_q^2\,\left [
C_A\,C_F^2\,{\overline{
|\Delta T^{(2)}_{q\bar q}|^2}}_{C_F} + C_A^2\,C_F\,
{\overline{ |\Delta T^{(2)}_{q \bar q}|^2}}_{C_A}\right ]\,,
\\[2ex]
\label{eqn3.8}
&&q + g \rightarrow q + g + \gamma^*\,,
\nonumber\\[2ex]
\overline{|\Delta M^{(2)}_{qg\rightarrow qg~\gamma^*}|^2}
&=&g^4\,e_q^2\,\left [C_A\,C_F^2\,{\overline{|\Delta T^{(2)}_{qg}|^2}}_{C_F}
+C_A^2\,C_F\,{\overline{|\Delta T^{(2)}_{qg}|^2}}_{C_A}\right ]\,.
\end{eqnarray}
The colour and spin average factors $K_{q\bar q}$ and
$K_{qg}$ are given in Eqs. (\ref{eqn2.25}) and (\ref{eqn2.26})
respectively. Further the indices $C_F$ and $C_A$ refer to the highest power of
the colour factor multiplying the corresponding matrix elements. Finally
we have also explicitly indicated the charge of the quark indicated by $e_q$.
For the lowest order matrix elements in section 2 we have implicitly taken it
into account. However as we will see below in higher order processes the 
Feynman graphs contain different species of quarks. Their charges will appear
in the corresponding coefficient functions and they indicate how they
have to be combined with the parton densities. 
In NLO we encounter some new subprocesses.
The first one is given by quark-quark
scattering (non-identical and identical quarks) represented by
\begin{eqnarray}
\label{eqn3.9}
&&q_1 + q_2 \rightarrow q_1 + q_2 + \gamma^*\,, \qquad q_1\not = q_2\,,
\nonumber\\[2ex]
\overline{|\Delta M^{(2)}_{q_1q_2\rightarrow q_1q_2~\gamma^*}|^2}&=&
g^4\,C_A\,C_F\,\left [e_{q_1}^2
\,{\overline{|\Delta T^{(2)}_{q_1q_2}|^2}}_C+e_{q_2}^2\,{\overline{
|\Delta T^{(2)}_{q_1q_2}|^2}}_D\right.
\nonumber\\[2ex] 
&&\left. +e_{q_1}\,e_{q_2}\,{\overline{|\Delta T^{(2)}_{q_1q_2}|^2}}_{CD}
\right ] \,,
\\[2ex]
\label{eqn3.10}
&& q + q \rightarrow q + q + \gamma^*\,,
\nonumber\\[2ex]
\overline{|\Delta M^{(2)}_{qq\rightarrow qq~\gamma^*}|^2}&=&g^4\,C_F\,e_q^2\,
\left [C_A\left \{
{\overline{|\Delta T^{(2)}_{q_1q_2}|^2}}_C+{\overline{
|\Delta T^{(2)}_{q_1q_2}|^2}}_D+
{\overline{|\Delta T^{(2)}_{q_1q_2}|^2}}_{CD}\right \}\right.
\nonumber\\[2ex] 
&&\left. +{\overline{|\Delta T^{(2)}_{qq}|^2}}_{CDEF}\right ] \,,
\end{eqnarray}
In Eqs. (\ref{eqn3.9}), (\ref{eqn3.10}) $C$, $D$, $E$ and $F$ 
refer to the graphs in Figs.  8, 9 in \cite{maha}.
The second subprocess is quark-anti-quark scattering
\begin{eqnarray}
\label{eqn3.11}
&&q_1 + \bar q_2 \rightarrow q_1 + \bar q_2 + \gamma^*\,, 
\qquad q_1\not = q_2\,,
\nonumber\\[2ex]
\overline{|\Delta M^{(2)}_{q_1\bar q_2\rightarrow q_1\bar q_2~\gamma^*}|^2}&=&
g^4\,C_A\,C_F\,\left [e_{q_1}^2
\,{\overline{|\Delta T^{(2)}_{q_1q_2}|^2}}_C+e_{q_2}^2\,{\overline{
|\Delta T^{(2)}_{q_1q_2}|^2}}_D\right.
\nonumber\\[2ex]
&&\left. -e_{q_1}\,e_{q_2}\,{\overline{|\Delta T^{(2)}_{q_1q_2}|^2}}_{CD}
\right ] \,,
\\[2ex]
\label{eqn3.12}
&&q + \bar q \rightarrow q_i + \bar q_i + \gamma^*\,,\qquad q_i\not =q \,,
\nonumber\\[2ex]
\overline{|\Delta M^{(2)}_{q\bar q\rightarrow q_i\bar q_i~\gamma^*}|^2}&=&
g^4\,C_A\,C_F\,\left [
e_q^2\,(n_f-1){\overline{ |\Delta T^{(2)}_{q\bar q}|^2}}_A+\sum_{i=1}^{n_f-1} 
e_{q_i}^2\,{\overline{ |\Delta T^{(2)}_{q\bar q}|^2}}_B\right ]\,,
\nonumber\\[2ex]
\\[2ex]
\label{eqn3.13}
&&q + \bar q \rightarrow q + \bar q + \gamma^*\,,
\nonumber\\[2ex]
\overline{|\Delta M^{(2)}_{q\bar q\rightarrow q\bar q~\gamma^*}|^2}&=&
g^4\,C_F\,e_q^2\,\left [C_A\left \{
{\overline{ |\Delta T^{(2)}_{q\bar q}|^2}}_A 
+{\overline{ |\Delta T^{(2)}_{q\bar q}|^2}}_B
+ {\overline{|\Delta T^{(2)}_{q_1q_2}|^2}}_C \right.\right.
\nonumber\\[2ex]
&&+\left.\left.{\overline{ |\Delta T^{(2)}_{q_1q_2}|^2}}_D- {\overline{|\Delta 
T^{(2)}_{q\bar q}|^2}}_{CD}\right \}
+{\overline{|\Delta T^{(2)}_{q\bar q}|^2}}_{\rm ABCD}\right ]\,,
\end{eqnarray}
and the colour average factors read
\begin{eqnarray}
\label{eqn3.14}
K_{q_1q_2}=K_{qq}= K_{q_1\bar q_2}=K_{q\bar q}=\frac{1}{N^2}\,.
\end{eqnarray}
In Eqs. (\ref{eqn3.11}), (\ref{eqn3.12}) and (\ref{eqn3.13})
$A$, $B$, $C$ and $D$ refer to the graphs in Figs.  7, 8 in \cite{maha}.
The last new process which shows up in NLO is given by
\begin{eqnarray}
\label{eqn3.15}
&&g + g \rightarrow q_i + \bar q_i + \gamma^*\,,
\nonumber\\[2ex]
\overline{|\Delta M^{(2)}_{gg\rightarrow q_i\bar q_i~\gamma^*}|^2}&=&
g^4\,\sum_{i=1}^{n_f}e_{q_i}^2\,
\left [ C_A\,C_F^2\,{\overline{
|\Delta T^{(2)}_{gg}|^2}}_{C_F} + C_A^2\,C_F\,
{\overline{ |\Delta T^{(2)}_{gg}|^2}}_{C_A}\right ]\,,
\nonumber\\[2ex]
\mbox{with}&& K_{gg}=\frac{1}{(N^2-1)^2}\,,
\end{eqnarray}
with the same notation as explained below Eq. (\ref{eqn3.8}).
If the matrix elements are computed in $n$ dimensions one has to be careful
with the interference terms indicated by the indices $CD$, $CDEF$ in 
Eq. (\ref{eqn3.10}) and $ABCD$ in Eq. (\ref{eqn3.13}). 
These terms do not lead to collinear
divergences so they can be computed in four dimensions. However if they
are computed in $n$ dimensions the prescription for the $\gamma_5$-matrix
leads to additional terms which are proportional $\varepsilon=n-4$. After
integration over the angles in Eq. (\ref{eqn3.6}) these terms contribute
because they are multiplied by pole terms $1/\varepsilon$. Since a finite
quantity cannot depend on the regularization these terms are considered
to be spurious and they should be discarded. Note that without the
$\gamma_5$-matrix the terms proportional to $\varepsilon$ cancel automatically 
among themselves.
Some of the expressions i.e. Eqs. (\ref{eqn3.9}), (\ref{eqn3.10}), 
(\ref{eqn3.12}), (\ref{eqn3.13})
are singular in the limit $s_4\rightarrow 0$, where $s_4$ is defined in
Eq. (\ref{eqn3.5}), which is due
to soft gluon radiation or soft (collinear) fermion pair production.
This will lead to infrared singularities when the differential cross section
in Eq. (\ref{eqn3.5}) is convoluted with the parton densities. Therefore
the cross section will be split up into a hard gluon ($s_4>\Delta$) part and 
a soft gluon ($s_4\le \Delta$) part (see e.g. \cite{rasm}). The latter is
defined by
\begin{eqnarray}
\label{eqn3.16}
s\,\frac{d^2~{\hat W}^{\rm SOFT}_{a_1a_2}}{d~t~d~u}&=&
\frac{1}{2\pi} K_{a_1a_2}\,\frac{S_{\varepsilon}^2}
{(4\pi)^4\,\Gamma(1+\varepsilon)}\,
\left (\frac{t~u}{\mu^2~s}\right )^{\varepsilon/2}
\nonumber\\[2ex]
&& \times\delta(s+t+u-Q^2)\,\int_0^{\Delta}ds_4\,
\left ( \frac{s_4}{\mu^2}\right)^{\varepsilon/2} 
\overline{|\Delta M^{(2)}_{a_1a_2\rightarrow b_1b_2~\gamma^*}|^2}\,.
\nonumber\\
\end{eqnarray}
Only the singular part of 
$\overline{|\Delta M^{(2)}_{a_1a_2\rightarrow b_1b_2~\gamma^*}|^2}$, 
which behaves as $1/s_4$, contributes to the above integral 
whereas the non-singular terms vanish in the limit $s_4\rightarrow 0$. 
The calculation of the soft gluon cross section has to proceed in the same
way as done for the virtual corrections in Eqs. (\ref{eqn3.3}), (\ref{eqn3.4}).
First we apply the eikonal approximation to the singular part of the
matrix elements and apply tensorial reduction to the soft gluon phase space
integrals before we contract the Levi-Civita tensors in four dimensions.
The results which hold in the HVBM- as well as RSN-schemes are given by
\begin{eqnarray}
\label{eqn3.17}
s\,\frac{d^2~{\Delta {\hat W}}^{\rm SOFT}_{q\bar q}}
{d~t~d~u}&=& \delta(s+t+u-Q^2)\,
S_{\varepsilon}^2\,a_s^2\,
\frac{1}{N}
\nonumber\\[2ex]
&&\times \Bigg [n_f\,C_F\,\Bigg \{-\frac{4}{3\varepsilon}-
\frac{2}{3}\,\ln \frac{\Delta}{\mu^2} -\frac{2}{3}\,\ln \frac{tu}{\mu^2 s}
+\frac{10}{9}\Bigg \}
\nonumber\\[2ex]
&&+ C_A\,C_F\,\Bigg \{-\frac{8}{\varepsilon^2}
-\left (8 \ln\frac{tu}{\mu^2 s}-\frac{22}{3}\right )\frac{1}{\varepsilon}
-4\,\ln\frac{\Delta}{\mu^2}\ln \frac{tu}{\mu^2 s}
\nonumber\\[2ex]
&&+2\,\ln^2\frac{\Delta}{\mu^2}-2\,\ln^2\frac{tu}{\mu^2 s}
+\frac{11}{3}\ln\frac{\Delta}{\mu^2}
+\frac{11}{3}\ln \frac{tu}{\mu^2 s}
\nonumber\\[2ex]
&& +6\,\zeta(2)-\frac{67}{9}\Bigg \}
\nonumber\\[2ex]
&&+C_F^2\,\Bigg \{-\frac{16}{\varepsilon^2}-\left (16 \ln\frac{\Delta}{\mu^2}
\right )\,\frac{1}{\varepsilon}-8\ln^2\frac{\Delta}{\mu^2}+4\,\zeta(2)
\Bigg \}\Bigg ]
\nonumber\\[2ex]
&&\times |\Delta T^{(1)}_{q\bar q}|^2\,.
\end{eqnarray}
We have added the soft gluon contributions from Eqs. (\ref{eqn3.9}), 
(\ref{eqn3.12}) and (\ref{eqn3.13}) into the above expression. 
Like in the case of the virtual contribution in Eq. (\ref{eqn3.3})
the expression above is, apart from an overall minus sign and the
$\varepsilon$-dependence of $|\Delta T^{(1)}_{q\bar q}|^2$, equal to
the unpolarized structure function. Note that this $\varepsilon$-dependence
leads to the breakdown of the relation in Eq. (\ref{eqn2.62}).
From the reaction in Eq. (\ref{eqn3.9}) we obtain
\begin{eqnarray}
\label{eqn3.18}
s\,\frac{d^2~{\Delta {\hat W}}^{\rm SOFT}_{qg}}{d~t~d~u}&=&
\delta(s+t+u-Q^2)\,
S_{\varepsilon}^2\,a_s^2\,
\frac{1}{N}
\nonumber\\[2ex]
&&\times \Bigg [C_A\,\Bigg \{\frac{4}{\varepsilon^2}
+\left (4 \ln\frac{\Delta}{\mu^2}\right )\frac{1}{\varepsilon}
+2\,\ln^2\frac{\Delta}{\mu^2}-\zeta(2) \Bigg \}
\nonumber\\[2ex]
&&+C_F\,\Bigg \{\frac{8}{\varepsilon^2}+\Big (4\ln\frac{\Delta}{\mu^2}
+4\ln \frac{tu}{\mu^2 s}-3\Big )\frac{1}{\varepsilon}
+2\,\ln\frac{\Delta}{\mu^2}\ln \frac{tu}{\mu^2 s}
\nonumber\\[2ex]
&&+\ln^2\frac{\Delta}{\mu^2}
+\ln^2\frac{tu}{\mu^2 s}-\frac{3}{2}\,\ln\frac{\Delta}{\mu^2}
-\frac{3}{2}\ln \frac{tu}{\mu^2 s}
\nonumber\\[2ex]
&&-4\zeta(2)+\frac{7}{2}\Bigg \}\Bigg ]\,
|\Delta T^{(1)}_{qg}|^2\,.
\end{eqnarray}
The infrared and final state collinear divergences are cancelled upon 
adding the virtual contributions in Eqs. (\ref{eqn3.3}) and (\ref{eqn3.4})
to the soft gluon contributions above. 
The ultraviolet divergences are removed by subtracting from 
$d^2~\Delta {\hat W}^{(2)}_{a_1a_2}/dt~du$ ($a_1,a_2=q,g$) the counter term 
\begin{eqnarray}
\label{eqn3.19}
 a_s^2(\mu^2)\,S_{\varepsilon}\,
\frac{2\beta_0}{\varepsilon}\,s 
\frac{d^2~{\Delta {\hat W}}^{(1)}_{a_1a_2}}{d~t~d~u}\,,\quad \mbox{with}\quad
\beta_0=\frac{11}{3}\,C_A-\frac{4}{3}\,T_f\,n_f \,, 
\end{eqnarray}
where we have chosen the ${\overline {\rm MS}}$ renormalization scheme.
Here $\beta_0$ denotes the lowest order contribution to the $\beta$-function.
After renormalization one still has to perform mass factorization to
remove the remaining collinear divergences. This is achieved by the formula
\begin{eqnarray}
\label{eqn3.20}
s\,\frac{d^2~{\Delta {\hat W}}_{a_1a_2}}{d~t~d~u}(s,t,u,
\varepsilon)
&=&\sum_{b_1,b_2=q,g}\int_0^1 \frac{dx_1}{x_1}\,\int_0^1 \frac{dx_2}{x_2}\,
\Delta \Gamma_{b_1a_1}(x_1,\varepsilon)\,\Delta \Gamma_{b_2a_2}(x_2,\varepsilon)
\nonumber\\[2ex]
&&\times {\hat s}\,\frac{d^2~\Delta W_{b_1b_2}}
{d~\hat t~d~\hat u} (\hat s,\hat t,\hat u)
\nonumber\\[2ex]
&&\equiv \sum_{b_1,b_2=q,g}\Delta \Gamma_{b_1a_1}\otimes\Delta \Gamma_{b_2a_2}
\otimes {\hat s}\, \frac{d^2~\Delta W_{b_1b_2}}{d~\hat t~d~\hat u}\,,
\end{eqnarray}
with the following definitions
\begin{eqnarray}
\label{eqn3.21}
\hat s = x_1\,x_2\,s\,, \qquad \hat t=x_1\,(t-Q^2)+Q^2\,,\qquad
\hat u=x_2\,(u-Q^2)+Q^2\,.
\end{eqnarray}
In Eq. (\ref{eqn3.20}) $d^2~{\Delta {\hat W}}_{a_1a_2}$ represents the singular 
structure functions which contain the collinear divergences indicated by 
$\varepsilon$.
These divergences are removed by the kernels $\Delta \Gamma_{b_ia_i}$ leaving 
the finite coefficient functions $d^2~\Delta W_{b_1b_2}$. For the RSN 
prescription
the kernels are given in Eqs. (\ref{eqn2.67})- (\ref{eqn2.70}). In the case of 
the off-shell mass assignment they are represented by the operator matrix 
elements in Eqs. (\ref{eqn2.45})- (\ref{eqn2.51}).
Both the kernels and the coefficient functions depend on the mass 
factorization scale $\mu$. Both regularization techniques lead to the
same coefficient functions renormalized in ${\overline {\rm MS}}$-scheme.
The coefficient functions originating from
the soft-plus-virtual gluon contributions are sufficiently short 
that they can be published and one can find them in Appendix A. 
The hard gluon parts are too long to be published.
They exist as FORM \cite{form} files and they are available on request.
Here we can only show those parts which behave like $1/s_4$.
In the soft gluon limit where $s_4 \rightarrow 0$ they read
\begin{eqnarray}
\label{eqn3.22}
 \mathop{\mbox{lim}}\limits_{\vphantom{\frac{A}{A}} s_4 \rightarrow 0}
s\,\frac{d^2~\Delta W^{\rm HARD}_{q \bar q}}{d~t~d~u}&=&
a_s(\mu^2)\,\frac{1}{N^2}\,
\frac{1}{s_4}\,\Bigg [ n_f\,T_f\,C_F\,\frac{4}{3} +
\nonumber\\[2ex]
&&+C_A\,\Bigg \{4\,\ln \frac{s_4}{\mu^2}
-4\,\ln \frac{tu}{\mu^2 s}+\frac{11}{3}\Bigg \}
\nonumber\\[2ex]
&& +C_F\,\Bigg \{ -16\,\ln \frac{s_4}{\mu^2}+8\ln \frac{tu}{\mu^2 s}
\Bigg \} \Bigg ]\, |\Delta M^{(1)}_{q\bar q}|^2\,,
\nonumber\\
\\[2ex]
\label{eqn3.23}
 \mathop{\mbox{lim}}\limits_{\vphantom{\frac{A}{A}} s_4 \rightarrow 0}
s\,\frac{d^2~\Delta W^{\rm HARD}_{q g}}{d~t~d~u}&=&
a_s(\mu^2)\,\frac{1}{N}\,
\frac{1}{s_4}\,\Bigg [C_A\,\Bigg \{4\ln \frac{s_4}{\mu^2}
-2\,\ln \frac{tu}{\mu^2 s}\Bigg \}
\nonumber\\[2ex]
&& +C_F\,\Bigg \{2\,\ln \frac{s_4}{\mu^2}
-\frac{3}{2} \Bigg \} \Bigg ] \, |\Delta M^{(1)}_{qg}|^2\,.
\end{eqnarray}
Another important result which emerges from our calculation is that
we find the same relation as in Eq. (\ref{eqn2.62}) 
for the non-singlet part of the coefficient function 
i.e.
\begin{eqnarray}
\label{eqn3.24}
s\,\frac{d^2~\Delta W_{q\bar q}^{\rm NS,(2)}}{d~t~d~u}
=-s\,\frac{d^2~W_{q\bar q}^{\rm NS,(2)}}{d~t~d~u}\,,
\end{eqnarray}
which holds for the hard and soft-plus-virtual part independently
(see also the comment below Eq. (\ref{eqnA.1}). The sum of these two parts
is in agreement with Eq. (60) in \cite{chco2} provided one omits the 
interference contribution called $2\sigma_4$ which originates 
from the non-singlet part of the $qq$-channel. The relation in 
Eq. (\ref{eqn3.24}) follows from 
chiral symmetry because the quarks are massless. Therefore the finite 
coefficient functions should respect this relation irrespective which 
regularization method is used.
Finally we want to comment on the scale $\mu$.
In the computation of the radiative corrections we have
assumed that the renormalization scale $\mu_r$ is equal to the mass
factorization scale $\mu$. If one wants to distinguish between these scales
one has to substitute
\begin{eqnarray}
\label{eqn3.25}
a_s(\mu^2)=a_s(\mu_r^2)\,\left (1+a_s(\mu_r^2)
\beta_0 \, \ln \frac{\mu_r^2}{\mu^2}\right)\,,
\end{eqnarray}
in all finite expressions.

\mysection{Differential distributions for the process\\
 $p + p\rightarrow \gamma^* +'X'$}
In this section we present the differential distributions in Eqs. 
(\ref{eqn2.3}), (\ref{eqn2.4})
for lepton-pair production in proton-proton collisions at the RHIC 
and make a comparison with similar results in previous work.
In practice one is not interested in the distributions which depend on
T and U in Eq. (\ref{eqn2.2}) but in the differential cross section
given by
\begin{eqnarray}
\label{eqn4.1}
\frac{d^3~\Delta \sigma^{{\rm pp}}}{d~Q~d~p_T~d~y}(S,p_T^2,y,Q^2)=4\,S\,Q\,p_T
\,\frac{d^3~\Delta \sigma^{{\rm pp}}}{dQ^2~d~T~d~U}(S,T,U,Q^2)\,,
\end{eqnarray}
where $y$ and $p_T$ denote the rapidity and transverse momentum respectively.
Neglecting the masses of the incoming hadrons we have the following relations
\begin{eqnarray}
\label{eqn4.2}
&&T=Q^2-\sqrt S\,\sqrt{p_T^2+Q^2}\,\cosh y+\sqrt S\,\sqrt{p_T^2+Q^2}
\,\sinh y\,,
\nonumber\\[2ex]
&&U=Q^2-\sqrt S\,\sqrt{p_T^2+Q^2}\,\cosh y-\sqrt S\,\sqrt{p_T^2+Q^2}
\,\sinh y\,.
\end{eqnarray}
The kinematical boundaries are
\begin{eqnarray}
\label{eqn4.3}
Q^2-S\le T \le 0 \,, \qquad -S-T+Q^2\le U \le \frac{S~Q^2}{T-Q^2}+Q^2\,,
\end{eqnarray}
from which one can derive
\begin{eqnarray}
\label{eqn4.4}
&& 0\le p_T^2 \le p^2_{T,{\rm max}}\,, \qquad
- \frac{1}{2}\ln \frac{S}{Q^2}\le y \le \frac{1}{2}\ln \frac{S}{Q^2}\,,
\nonumber\\[2ex]
&&\mbox{with} \quad p^2_{T,{\rm max}}=\frac{(S+Q^2)^2}{4~S~\cosh^2 y}-Q^2\,,
\end{eqnarray}
or
\begin{eqnarray}
\label{eqn4.5}
&&- y_{{\rm max}}\le y \le y_{{\rm max}}\,,
\qquad 0\le p_T^2 \le \frac{(S-Q^2)^2}{4~S}\equiv {\bar p}^2_{T,{\rm max}} \,,
\nonumber\\[2ex]
&&\mbox{with} \quad y_{{\rm max}}=
\frac{1}{2}\ln\frac{1+\sqrt{1-sq}}{1-\sqrt{1-sq}}\,,
\qquad sq=\frac{4~S~(p_T^2+Q^2)}{(S+Q^2)^2}\,.
\end{eqnarray}
Since the cross section in Eq. (\ref{eqn4.1}) diverges for $p_T\rightarrow 0$ 
we cannot perform
the integral over this kinematical variable down to zero. However the
full integration over the rapidity can be carried out and one obtains
the transverse momentum distribution
\begin{eqnarray}
\label{eqn4.6}
\frac{d^2~\Delta \sigma^{{\rm pp}}}{d~Q~d~p_T}(S,p_T^2,Q^2)=
\int_{-y_{{\rm max}}}^{y_{{\rm max}}} dy
\,\frac{d^3~\Delta \sigma^{{\rm pp}}}{d~Q~d~p_T~d~y}
(S,p_T^2,y,Q^2)\,,
\end{eqnarray}
with $y_{{\rm max}}$ given in Eq. (\ref{eqn4.5}).
Finally we define what we mean by leading order (LO) and next-to-leading
order (NLO). In LO the differential cross section is determined by
the leading logarithmic approximations to the coupling constant 
$\alpha_s(\mu_r^2)$ in Eq. (\ref{eqn3.25})
and the polarized parton densities $\Delta f_{a_i}^P(x,\mu^2)$ in 
Eq. (\ref{eqn2.4}). In NLO the latter two quantities are replaced by
their next-to-leading logarithmic approximations.

In the subsequent part of this section we will study the dependence of the
cross sections as defined in Eqs. (\ref{eqn4.1}), (\ref{eqn4.6}) on input
parameters like the QCD scale $\Lambda$, the factorization
scale $\mu$ and the dependence on the chosen set of polarized parton densities. 
Notice that we have adopted $\mu_r=\mu$ for the renormalization scale. 
In our computations the number of light flavours is taken
to be $n_f=4$ which holds for the running coupling,
the DY coefficient functions and the polarized parton densities.
Further we have chosen for our plots the polarized parton densities
provided by GRSV \cite{grsv} (here called GRSV01) and BB \cite{blbo}, 
which were determined in the $\overline {\rm MS}$ renormalization scheme.
Notice that these sets do not contain charm and bottom quark densities.
The densities of GRSV standard scenario were constructed to fit
the available data together with positivity requirements
and to satisfy two SU(3) flavour group sum rules.
The GRSV collaboration have also given another set of parton densities,
the so-called valence set, where the sum rules hold for the SU(2)
flavour group as suggested originally in \cite{lipkin}.
It turns out that in the standard scenario the gluon density is larger
than in the valence scenario so that the DY cross section will be dominated by 
the $qg$ process in the former scenario provided the transverse momentum will 
be sufficiently large. However in the valence scenario the sea-quark
density becomes important too. Therefore there will be a competition
between the $qg$ 
\footnote{ Note that the notation $qg$ represents the subprocesses $qg$ and
$\bar q g$. The same holds for the notation $qq$ representing the reactions
$qq$ and $\bar q \bar q$.}
and $q\bar q$ subprocesses and it will be very hard to disentangle
these two densities.
The BB collaboration have presented two analyses of polarized deep
inelastic scattering data. At small $x$ the densities behave as 
$x^a$ where the $a_G$ for the gluon and the $a_S$ for the sea quark
are related by $a_G = a_S + c$. In the first scenario BB assume
$c=0.9$ in LO (ISET=1) and $c=1.0$ in NLO (ISET=3).
In the second scenario BB assume $c=0.6$ in LO (ISET=2) and $c=0.5$
in NLO (ISET=4). All sets above are presented in LO and NLO with the
$\Lambda_4$ and the corresponding values for $\alpha_s(M_Z)$ in
Table 1.
Besides the polarized cross sections $d\Delta \sigma$ we also want to compute
the double longitudinal spin asymmetry defined by
\begin{eqnarray}
\label{eqn4.7}
A_{LL} = \frac{d\Delta \sigma}{d\sigma} \,,
\end{eqnarray}
where $d\sigma$ denotes the unpolarized cross section which is calculated
up to NLO in \cite{arre}. \footnote{We re-calculated the unpolarized DY
cross section and found agreement with \cite{arre}. However we used 
our versions of the coefficient functions for the plots in this paper.}
In order to compute
the latter we adopt the GRV \cite{grv98} parton density set 
(here called GRV98) which does not contain any
charm or bottom quark densities either (for more details see Table. 1).
Before presenting the results we note that our computer programs were checked
by reproducing Fig.4 in \cite{chco2} for the LO polarized and unpolarized
nonsinglet $p_T$-distributions. The difference
between their results and ours is numerically small which might be due to
a different choice of $\alpha_s$. Notice that the interference term
$2\sigma_4$ mentioned below Eq. (\ref{eqn3.24}) is completely negligible. 
Also we reproduced
several results in \cite{bego2} including in particular Fig.3 for the
unpolarized NLO inclusive $p_T$ distribution and Fig. 8(b) for the LO 
contribution to $A_{\rm LL}(p_T)$ in Eq. (\ref{eqn4.7}).
\begin{table}
\begin{center}
\begin{tabular}{|c|c|c|}\hline
GRSV01 (LO, standard scenario)& $\Lambda_4^{\rm LO}=175~{\rm MeV}$  &
$\alpha_s^{\rm LO}(M_Z)=0.121$  \\
GRSV01 (NLO, standard scenario)  & $\Lambda_4^{\rm NLO}=257~{\rm MeV}$ &
$\alpha_s^{\rm NLO}(M_Z)=0.109$       \\
GRSV01 (LO, valence scenario)     & $\Lambda_4^{\rm LO}=175~{\rm MeV}$   &
$\alpha_s^{\rm LO}(M_Z)=0.121$       \\
GRSV01 (NLO, valence scenario)    & $\Lambda_4^{\rm NLO}=257~{\rm MeV}$  &
$\alpha_s^{\rm NLO}(M_Z)=0.109$    \\
BB (LO, scenario 1) & $\Lambda_4^{\rm LO}=203~{\rm MeV}$   &
$\alpha_s^{\rm LO}(M_Z)=0.123$   \\
BB (NLO, scenario 1) & $\Lambda_4^{\rm NLO}=235~{\rm MeV}$  &
$\alpha_s^{\rm NLO}(M_Z)=0.107$       \\
BB (LO, scenario 2) & $\Lambda_4^{\rm LO}=195~{\rm MeV}$  &
$\alpha_s^{\rm NLO}(M_Z)=0.123$       \\
BB (NLO, scenario 2) & $\Lambda_4^{\rm NLO}=240~{\rm MeV}$ &
$\alpha_s^{\rm NLO}(M_Z)=0.107$ \\
GRV98 (LO) & $\Lambda_4^{\rm LO}=175~{\rm MeV}$ &
$\alpha_s^{\rm NLO}(M_Z)=0.121$ \\
GRV98(NLO) & $\Lambda_4^{\rm NLO}=257~{\rm MeV}$ &
$\alpha_s^{\rm NLO}(M_Z)=0.109$ \\
\hline
\end{tabular}
\end{center}
\caption{Polarized and unpolarized parton density sets with the values for 
the QCD scale $\Lambda_4$ and the running coupling $\alpha_s(M_Z)$.}
\label{table1}
\end{table}

For our plots we choose $\sqrt S = 200~{\rm GeV}$ which is a representative 
C.M. energy for proton-proton collisions at the RHIC. The factorization
scale is set $\mu^2=Q^2+p_T^2$ unless mentioned otherwise.
We begin with Figs. 1a, 1b which show the $p_T$ dependence in LO and NLO
of Eq. (\ref{eqn4.6}) for $Q = 6~{\rm GeV/c}$ with the set GRSV01 (standard 
scenario). Both in LO and NLO the $qg$ subprocess is positive but the 
$q\bar q$ channel yields 
larger contributions than the $qg$ channel when $p_T<3~{\rm GeV/c}$.
For larger transverse momenta the $qg$ subprocess dominates 
and the $q\bar q$ contribution drops off rapidly where it even
becomes negative when $p_T >27~{\rm GeV/c}$. 
At $p_T \approx 6~{\rm GeV/c}$ the cross section due to the $q\bar q$ 
subprocess is only one quarter of the contribution of the $qg$ channel. 
Hence we see that when $p_T > Q/2$ the $qg$ channel begins to 
dominate over the other subprocesses.
The NLO results from all subprocesses are presented in Fig.1b 
where we note that the contributions from 
the $gg$ and $qq$ channels are small and negative so that we plot 
their absolute values. The possibility that the 
polarized gluon density can be measured in the DY process at large $p_T$ 
has already been stressed by several authors using LO perturbation 
expressions, for recent work see \cite{bego2}.
We have now demonstrated that this conclusion is unaltered 
when the NLO contributions are included.

In Figs. 2a, 2b we do the same as in Figs. 1a ,1b but now the plots are 
presented for the GRSV01 valence scenario. In this case the $q\bar q$ channel 
also dominates the cross section at small transverse momenta but 
this contribution is negative over the whole $p_T$ range
so we plot its absolute value.
On the other hand the $qg$ subprocess only yields positive contributions.
The net effect is that below $p_T\sim 5~{\rm GeV/c}$ the LO polarized 
cross section due to  both
contributions is negative because the $q\bar q$ reaction is dominant in
this region. For  $p_T> 5~{\rm GeV/c}$ the LO cross section becomes positive 
since in this region the $qg$-channel is more important. Hence we plot the
absolute value of LO(sum) in Fig. 2a.
This overall picture is not altered by including the NLO 
corrections so also the NLO(sum) is negative for small $p_T$ and
positive for large $p_T$ and we add the plot of its absolute value to Fig.2a. 
In Fig. 2b we show the individual NLO contributions. The $gg$ and $qq$ 
channels are small and negative so we have plotted their absolute values.
Therefore the main difference between the standard and valence scenarios 
shows up in the region $p_T<5~{\rm GeV/c}$ where the standard scenario 
provides us with a positive NLO distribution while the valence scenario 
yields a negative one. Note that there is nothing unphysical about this
result because the polarized cross section is the difference between
two helicity projections.  If the valence scenario is correct one will 
need high statistics in the region where 
the $p_T$ distribution changes sign, say for $4<p_T<6~{\rm GeV/c}$.
 
Given the changes in signs in the above results it is clear that rapidity
plots integrated over a range in $p_T$ will depend strongly on the 
range available experimentally and will also vary dramatically . 
Therefore we will only present the rapidity distributions 
in Eq. (\ref{eqn4.1}), at a fixed representative $p_T$. 
In Figs. 3a, 3b we show the LO and NLO corrected distributions 
with $Q=6~{\rm GeV/c}$ and $p_T=2~{\rm GeV/c}$ using the GRSV01 standard
scenario parton densities. In this region the $q\bar q$
channel dominates the cross section and it leads to a positive contribution
in the whole rapidity range. Since the $qg$ channel has two maxima at 
symmetric values of $y$ the total LO distribution has a slight minimum at
$y=0$ which is enhanced when the NLO corrections are taken into account.
This is explained by Fig. 3b where we see that the $qg$, $gg$ and $qq$ 
contributions are all negative near $y=0$ so the dip is more pronounced. 
At larger $p_T$ the relative contributions from these
channels changes but the $qg$ becomes a larger fraction of the total so
that the double peaked nature of the NLO distribution will remain.
We repeat these two plots for the valence 
scenario parton densities in Figs. 4a and 4b.
However we have now chosen $p_T=6~{\rm GeV/c}$ where the
both the LO and NLO cross sections are positive. 
Here one clearly sees that the rapidity distributions are negative
near $y=0$, where they have pronounced minima. This is wholly due
to the $q\bar q$ subprocess which yields negative contributions to
the cross section over the whole rapidity range. The $qg$ subprocess 
is positive for all values of the rapidity and compensates the
$q\bar q$ contribution leaving a much smaller cross section than
shown by the standard scenario in Figs. 3a, 3b. This was already
expected from the transverse momentum distributions in Figs. 2a,2b 
where the cross section decreases at small values of $p_T$ and
is actually negative for $p_T<5~{\rm GeV/c}$. At larger $p_T$ the
double peaked nature of this distribution will prevail as the
relative contribution from the $qg$ channel increases. 
Note that the integrals over the above rapidity distributions agree 
with the values of the $p_T$ distributions in Figs. 1a, 1b  and Figs. 2a, 2b
at $p_T=2$ and $p_T=6$ GeV/c respectively.  

In Fig. 5 we have shown the LO and NLO corrected cross sections 
for the GRSV01 standard scenario at different
values for the di-lepton pair invariant mass $Q$. When we increase the value 
of $Q$ the $p_T$ distributions decrease in LO as well as in NLO.
Further we observe that the NLO corrections
become larger when the transverse momentum increases. This will become more
clear when we show the $K$-factor later on. The dependence on $Q$ is also 
investigated for the $y$-distributions shown in Fig. 6. Here the the 
NLO correction is conspicuous at $Q=2~{\rm GeV/c}$ whereas at larger values
of $Q$ there are hardly any differences between the LO and NLO corrected
cross sections.

Radiative corrections are marred by theoretical uncertainties. The first
one concerns the choice of the factorization and renormalization scale
$\mu$. Since the perturbation series is truncated at a certain order
of $\alpha_s$ physical quantities depend on $\mu$. Notice that this
dependence will disappear when all orders are taken into account.
The sensitivity to the scale above is exhibited by the ratio
\begin{eqnarray}
\label{eqn4.8}
N\left (\frac{\mu}{\mu_0}\right ) = 
\frac{d\Delta \sigma (\mu)} 
{d\Delta \sigma (\mu_0)} \,.
\end{eqnarray}
If this ratio is close to unity and almost independent of $\mu$ the
higher order corrected cross section will be very reliable.
In Fig. 7a we use the GRSV01 standard scenario and investigate the ratio
in Eq. (\ref{eqn4.8}) for the cross section in Eq. (\ref{eqn4.6}).
Here we choose as central value
$\mu_0^2=Q^2+ p_T^2$ and we vary $\mu$ from 0.2 $\mu_0$ to
5$\mu_0$ at $Q=8~{\rm GeV/c}$. 
Note that the scale on the $\mu/\mu_0$-axis is logarithmic. 
For the transverse momenta we take $p_T=2$, 10 and 20 ${\rm GeV/c}$. 
In Fig. 7a the LO corrected $N(p_T,\mu/\mu_0)$ are the upper 
curves for small $\mu/\mu_0$ whereas the lower
curves are the NLO corrected $N(p_T,\mu/\mu_0)$. 
One sees that the scale dependence diminishes in going from LO to NLO
which indicates better predictive power in NLO perturbation theory.  
The analogous plot is shown in Fig. 7b for 
$d^3\Delta\sigma(p_t,y,\mu)/dQdp_Tdy$ as a function of $y$,
with $y=0$ and $y=\pm 1$ where now $\mu_0^2=Q^2+ p_{T}^2$ and $p_{T}=2$ GeV/c. 
Since the $y$-plots are symmetric the values at $y=1$ are 
identical to those for $y=-1$ so there are only four curves in Fig. 7b. 
Again the LO curves are above the NLO ones for small $\mu/\mu_0$. Like
in the case of the transverse momentum distributions there is an improvement
in the scale dependence in going from LO to NLO perturbation theory.

Besides the dependence on the factorization/renormalization scales
there are two other uncertainties which affect the predictive power of the
theoretical cross section. The first one concerns the rate of convergence
of the perturbation series which is indicated by the $K$-factor defined by
\begin{eqnarray}
\label{eqn4.9}
K = \frac{d \Delta \sigma^{\rm NLO}}{d \Delta \sigma^{\rm LO}}.
\end{eqnarray}
This quantity is plotted as a function of $p_T$ in Fig. 8a for
$Q=2,6$ and $10~{\rm GeV/c}$ in the GRSV standard scenario
and we see that it varies from approximately
1.2 at $p_T = 2~{\rm GeV/c}$ to 1.7 at $p_T=30~{\rm GeV/c}$.
The corresponding rapidity plots for fixed $p_T=2$ GeV/c in Fig. 8b vary from 
0.8 to 1.5 for the case of $Q=2~{\rm GeV/c}$
and from 1.2 to 1.5 for the case of $Q=6~{\rm GeV/c}$. Notice the dips
around $y=0$. They are explained in the discussion of Figs. 3a,3b. The
minimum in the LO rapidity distribution becomes deeper when the NLO
corrections, mainly coming from the $qg$ subprocess, are included. 
The study of the $K$-factors reveal that
the NLO corrections are appreciable (sometimes more than $50\%$)
except at small $p_T$ where fortunately the cross section attains
its maximum. 

A third uncertainty which has to be solved by experiment are the
parton densities originating from different parametrizations. To study
the dependence on these densities it is better to reduce
the effect of the choice of factorization scale and the
various $K$-factors. This is achieved by plotting the double longitudinal 
spin asymmetry as given by Eq. (\ref{eqn4.7}). Besides the GRSV01 parton
density sets we also include the sets presented in \cite{blbo}. 

We show the LO double longitudinal spin asymmetry versus $p_T$ 
in Fig. 9a and the corresponding NLO result in Fig. 9b 
for $Q=6~{\rm GeV/c}$ and $\sqrt S= 200~{\rm GeV}$. They are integrated over
$y$ and the results are given in percent. The BB set 1 shows the most dramatic
rise in $A_{LL}$ when $p_T$ increases, while the other scenarios 
only show moderate increases. This is because the polarized gluon
density in this set is larger than those contained in the other sets.
Note that in the GRSV01 valence scenario $A_{LL}$ is negative in LO and NLO
below $p_T\sim 5~{\rm GeV/c}$ which can be inferred from Figs. 2a and 2b. 
However if the value of $A_{\rm LL}(p_T)$ is as small as indicated here 
this effect will not be measurable. 
A more direct comparison of the NLO results for $A_{LL}(p_T)$ in Fig. 9b 
with the LO results in Fig. 9a is provided by the ratios of the two plots 
which is given in Fig.9c. The strange behaviour of the GRSV01 valence
scenario below $p_T= 6~{\rm GeV/c}$ is due to the change in signs in the
LO and NLO results.
We see that this ratio is close to unity for
the GRSV standard scenario and the BB, set 1 while it is smaller for the
BB, set 2. This reflects the fact that the $K$-factors in the polarized 
and unpolarized $p_T$ distributions are almost equal. 
In Figs. 10a and 10b we show the LO and NLO results for $A_{\rm LL}(p_T,y)$
as a function of $y$ at $Q=6~{\rm GeV/c}$ and $p_{T}=6~{\rm GeV/c}$. 
For these values the $qg$ contribution dominates and the cross section 
is positive. Since the cross sections for the GRSV01 valence scenario 
in Figs. 4a,4b are small at this $p_T$ and fluctuate in sign across
the rapidity range it is likely that $A_{\rm LL}(p_T,y)$ will not be large 
as a function of $y$ so we have omitted this parametrization in Figs. 10a-c.
Again we see that the BB set 1 distributions yield the largest results.
We also take the ratios of the plots in Figs. 10b divided by those
in Figs. 10a, which are presented in Fig. 10c. 
The latter reveals that the ratio between
the LO and NLO corrected longitudinal asymmetry deviates considerably from
unity for the BB sets in particular for set 2. This means that the $K$-factors
for these sets differ from those computed for the unpolarized cross sections.
In the central rapidity
region the results are below unity reflecting the negative NLO contributions
found earlier. From Figs. 9b, 10b we can conclude that one can only
distinguish between the various parton densities when the polarized
gluon density is very large. In the case of the GRSV-set one cannot
measure the difference between the valence and the standard scenario.
Even the difference between the GRSV scenarios and scenario 2
of the BB-set is very small. If we assume that the proton beams have
100 $\%$ polarization ($75\%$ is more likely) we need to know
the polarized cross sections up to $7\%$ to obtain $A_{LL}$ in Fig. 9b at 
$p_T=20~{\rm GeV/c}$ up to $12.5\%$. This is necessary to distinguish
between the GRSV set and BB, set 2 which is very unlikely. However
if the error on the polarized cross sections is $17\%$ then $A_{LL}$ 
at $p_T=20~{\rm GeV/c}$ can be determined up to $25\%$ which is sufficient to 
distinguish between scenarios 1 and 2 of the BB-set.

Summarizing the above we have calculated the NLO corrections to the
single particle inclusive distributions for lepton-pair production
in polarized proton-proton scattering. We used
the approximation that the lepton pair emerges from a virtual photon only
and we neglected the additional contributions coming from the $Z$-boson 
production which is quite valid as long as 
$Q < 50~{\rm GeV/c}$. A part of the calculation was performed using two
different regularization methods to deal with the spurious terms introduced
by the prescription for the $\gamma_5$-matrix if the matrix elements are
computed in $n$ dimensions. The coefficient functions are presented in
the standard ${\overline {\rm MS}}$ scheme. In this scheme
the non-singlet polarized and unpolarized coefficient functions
satisfy the relation $\Delta W_{q \bar q} = - W_{q \bar q}$
which are strong checks on our calculations. We have 
presented some NLO results for quantities of experimental interest.
The fact that the $qg$ channel yields the largest contribution
at moderate $p_T$ indicates that experimental measurements of the
DY $p_T$ distribution will give us information on the polarized gluon density
provided the latter is sufficiently large. 
We found that the $K$ factors are moderate at least in the 
small $p_T$-range and that there is a reduction in the 
scale dependence of the inclusive distributions in NLO as compared with
LO. It is now up to the experimental groups working on polarized 
proton-proton scattering at RHIC to measure these DY distributions.

Acknowledgement: We thank M. Klasen for discussions regarding the plots
in \cite{bego2} and C. Coriano for a clarification of the results
in \cite{chco2}.


\appendix
\mysection*{Appendix A}
\setcounter{section}{1}
Here we list the soft-plus-virtual gluon contributions to the 
polarized DY coefficient functions which are determined by the 
one-loop corrections to the processes
in Eqs. (\ref{eqn2.25}), (\ref{eqn2.26}) and the soft gluon corrections
due to reactions (\ref{eqn3.7}), (\ref{eqn3.9}).
The $q\bar q$ reaction yields the result
\begin{eqnarray}
\label{eqnA.1}
s \frac{d^2~\Delta {\hat W}^{\rm S+V}_{q\bar q}}
{d~t~d~u}&=& \delta(s+t+u-Q^2)\,
S_{\varepsilon}^2\,a_s^2\, \frac{1}{N}
\nonumber\\[2ex]
&&\times \Bigg [n_f~C_F\,\Bigg \{
-\frac{2}{3}\ln\frac{\Delta}{\mu^2}+\frac{10}{9} \Bigg \}
|\Delta T^{(1)}_{q\bar q}|^2
\nonumber\\[2ex]
&&+C_F(C_F-C_A/2)\,\Bigg \{-4\,\ln^2 \frac{\Delta}{\mu^2}
+4\,\ln \frac{tu}{\mu^2 s} \ln \frac{\Delta}{\mu^2}
\nonumber\\[2ex]
&&-4\,{\rm Li}_2\left (\frac{s-Q^2}{s} \right )
-4\,{\rm Li}_2 \left (\frac{t}{Q^2}\right )
-4\,\ln^2\frac{-t}{\mu^2}-2\,\ln^2\frac{s}{\mu^2}
\nonumber\\[2ex]
&&-2\,\ln \frac{-t}{\mu^2}\ln \frac{-u}{\mu^2}+8\,\ln \frac{-t}{\mu^2}
\ln \frac{s}{\mu^2} +2\,\ln^2\left (\frac{t-Q^2}{t}\right )
\nonumber\\[2ex]
&&-2\,\ln^2\left( \frac{Q^2-t}{Q^2}\right )-3\,\ln \frac{Q^2}{\mu^2}
-6\,\zeta(2)+8 +\Big (t \Leftrightarrow u\Big ) \Bigg \}
\nonumber\\[2ex]
&&\times |\Delta T^{(1)}_{q\bar q}|^2
\nonumber\\[2ex]
&&+C_AC_F\,\Bigg \{-\ln^2 \frac{\Delta}{\mu^2}+\frac{11}{6}\,
\ln \frac{\Delta}{\mu^2}-2\,{\rm Li}_2 \left (\frac{t}{Q^2}\right ) 
\nonumber\\[2ex]
&&+\ln \frac{-t}{\mu^2}\ln \frac{-u}{\mu^2}
+\ln^2\left (\frac{t-Q^2}{t}\right )-\ln^2\left( \frac{Q^2-t}{Q^2}\right )
-\frac{3}{2}\,\ln \frac{Q^2}{\mu^2}
\nonumber\\[2ex]
&&-3\,\zeta(2)+\frac{5}{18} +\Big (t \Leftrightarrow u\Big )  \Bigg \} 
|\Delta T^{(1)}_{q\bar q}|^2
\nonumber\\[2ex]
&&+C_F(C_F-C_A/2)\,\Bigg \{\frac{16s+8t}{u}\,\Bigg ({\rm Li}_2 \left 
(\frac{t}{Q^2} \right ) + {\rm Li}_2\left (\frac{s-Q^2}{s} \right )
\nonumber\\[2ex]
&&+\frac{1}{2}\,\ln^2\frac{-t}{s}-\frac{1}{2}\,\ln^2\left (\frac{t-Q^2}{t}
\right )+\frac{1}{2}\,\ln^2\left( \frac{Q^2-t}{Q^2}\right ) \Bigg )
\nonumber\\[2ex]
&&+\frac{8s(s-2Q^2)}{(s-Q^2)^2}\,\ln\frac{s}{Q^2}
+\Bigg (\frac{4tu}{(Q^2-t)^2} 
-\frac{8(t-2u)}{Q^2-t}-16\Bigg )
\nonumber\\[2ex]
&&\times  \ln \frac{-t}{Q^2}+\frac{4(s+u)}{t}+\frac{8s}{s-Q^2}
+\frac{4u}{Q^2-t}-4 +\Big (t \Leftrightarrow u\Big )\Bigg \}
\nonumber\\[2ex]
&& +C_AC_F\,\Bigg \{\Bigg ( \frac{2tu}{(Q^2-t)^2}+\frac{8(u-t)}{Q^2-t}-8\Bigg )
\ln \frac{-t}{Q^2}-\frac{2(s+u)}{t}
\nonumber\\[2ex]
&&+\frac{2u}{Q^2-t}-2 +\Big (t \Leftrightarrow u\Big )\Bigg \} 
 \Bigg ]\,,
\end{eqnarray}
which, apart from an overall minus sign, is equal to the 
unpolarized soft-plus-virtual gluon contributions to the
unpolarized DY coefficient function. The $qg$ subprocess provides us with
the result
\begin{eqnarray}
\label{eqnA.2}
s \frac{d^2~\Delta{\hat W}^{\rm S+V}_{qg}}
{d~t~d~u}&=& \delta(s+t+u-Q^2)\,
S_{\varepsilon}^2\,a_s^2\,
\frac{1}{N}
\nonumber\\[2ex]
&&\times \Bigg [C_F\,\Bigg \{
\ln^2\frac{\Delta}{\mu^2}-\frac{3}{2}\,\ln\frac{\Delta}{\mu^2}
+2\,{\rm Li}_2 \left (\frac{t}{Q^2} \right )+2\,{\rm Li}_2 \left (\frac{u}{Q^2}
\right )
\nonumber\\[2ex]
&&-2\,\ln \frac{-t}{\mu^2}\ln \frac{-u}{\mu^2}+\ln^2 \frac{-t}{\mu^2}
-\ln^2\left (\frac{t-Q^2}{t}\right )
\nonumber\\[2ex]
&&-\ln^2\left (\frac{u-Q^2}{u}\right )
+\ln^2\left( \frac{Q^2-t}{Q^2}\right )+\ln^2\left( \frac{Q^2-u}{Q^2}\right )
\nonumber\\[2ex]
&&+3\,\ln \frac{Q^2}{\mu^2}+8\,\zeta(2)-\frac{11}{2} \Bigg \} 
|\Delta T^{(1)}_{qg}|^2
\nonumber\\[2ex]
&&+C_A\,\Bigg \{ 2\,\ln^2\frac{\Delta}{\mu^2}-2\,\ln \frac{tu}{\mu^2 s} 
\ln \frac{\Delta}{\mu^2} +{\rm Li}_2\left (\frac{s-Q^2}{s} \right )
\nonumber\\[2ex]
&&-{\rm Li}_2 \left (\frac{u}{Q^2}\right )
+2\,\ln\frac{-t}{\mu^2}\ln\frac{-u}{s}+\frac{1}{2}\,\ln^2 \frac{u}{s}
+\frac{1}{2}\,\ln^2\left (\frac{u-Q^2}{u}\right )
\nonumber\\[2ex]
&&-\frac{1}{2}\,\ln^2\left( \frac{Q^2-u}{Q^2}\right )
-2\,\zeta(2)+1 \Bigg \} |\Delta T^{(1)}_{qg}|^2
\nonumber\\[2ex]
&&+C_F\,\Bigg \{ \frac{4(t+2u)}{s}\,\Bigg (-{\rm Li}_2
\left (\frac{s-Q^2}{s} \right )+{\rm Li}_2 \left (\frac{t}{Q^2}\right )
\nonumber\\[2ex]
&&-\ln\frac{-u}{\mu^2} \ln\frac{-t}{s}+\frac{1}{2}\,\ln^2 \frac{-t}{\mu^2}
-\frac{1}{2}\,\ln^2 \frac{s}{\mu^2}-\frac{1}{2}\,\ln^2\left 
(\frac{t-Q^2}{t}\right )
\nonumber\\[2ex]
&& +\frac{1}{2}\,\ln^2\left( \frac{Q^2-t}{Q^2}\right )+5\,\zeta(2)\Bigg )
+\Bigg (\frac{2su}{(s-Q^2)^2} -\frac{2(s+4u)}{s-Q^2}\Bigg )
\nonumber\\[2ex]
&& \times \ln \frac{s}{Q^2}
+\Bigg (\frac{2st}{(Q^2-t)^2}+\frac{4(2s-t)}{Q^2-t}-8\Bigg )\ln \frac{-t}{Q^2}
-\frac{2u}{s}+\frac{2u}{t}
\nonumber\\[2ex]
&&+\frac{2s}{Q^2-t}-\frac{2u}{s-Q^2}-2 \Bigg \}
\nonumber\\[2ex]
&&+C_A\,\Bigg \{ \frac{2(t+2u)}{s}\,\Bigg ({\rm Li}_2\left (\frac{s-Q^2}{s}
\right )-{\rm Li}_2 \left (\frac{t}{Q^2}\right )
\nonumber\\[2ex]
&&+\ln\frac{-u}{\mu^2} \ln\frac{-t}{s}-\frac{1}{2}\,\ln^2 \frac{-t}{\mu^2}
+\frac{1}{2}\,\ln^2 \frac{s}{\mu^2}+\frac{1}{2}\,\ln^2\left
(\frac{t-Q^2}{t}\right )
\nonumber\\[2ex]
&&-\frac{1}{2}\,\ln^2\left( \frac{Q^2-t}{Q^2}\right )-5\,\zeta(2)\Bigg )
-\frac{2s}{s-Q^2} \ln \frac{s}{Q^2} 
\nonumber\\[2ex]
&& -\frac{2t}{Q^2-t}\ln \frac{-t}{Q^2}+\frac{2u}{s}-\frac{2u}{t}\Bigg \}
\Bigg ]\,.
\end{eqnarray}

%

\centerline{\bf \large{Figure Captions}}
%
\begin{description}
\item[Fig. 1a.]
The LO differential cross section $d^2\Delta \sigma/dQ~dp_T$
at $\sqrt{s}=200$ GeV for the GRSV01 standard scenario
plotted in the range $2<p_T<30~{\rm GeV/c}$
with $Q=6~{\rm GeV/c^2}$ and $\mu^2=Q^2+p_T^2$. 
The LO plots are for the subprocesses $q\bar q$ (dot-dashed line),
$qg$ (dashed line) and the sum (dotted line).
For convenience we also add the NLO result (solid line) from Fig. 1b. 
\item[Fig. 1b.]
Same as Fig. 1a but now for NLO. Further we have shown the contributions
from all subprocesses $q\bar q$ (dot-dashed line), $qg$ (long-dashed line),
$gg$ (dotted line) and $qq$ (short-dashed line). The latter
two reactions yield negative contributions we plot their absolute values.
\item[Fig. 2a.]
Same as in Fig. 1a but now using GRSV01 valence scenario. Since the 
$q\bar q$ contribution is negative at small $p_T$ we plot its absolute value.
For convenience we also add the NLO result (solid line) from Fig. 2b. 
\item[Fig. 2b.]
Same as in Fig. 1b but now using GRSV01 valence scenario. Again we plot
absolute values when the contributions are negative.
\item[Fig. 3a.]
The LO differential cross section $d^3\Delta \sigma/dQ~dp_T~dy$  
at $\sqrt{s}=200$ GeV for the GRSV01 standard scenario
with $Q=6~{\rm GeV/c^2}$, $p_T=2~{\rm GeV/c}$ and $\mu^2=Q^2+p_T^2$.
The LO plots are for the subprocesses $q\bar q$ (dot-dashed line),
$qg$ (dashed line) and the sum (dotted line).
For convenience we also add the NLO result (solid line) from Fig. 3b. 
\item[Fig. 3b.]
Same as Fig. 3a but now for NLO. Further we have shown the contributions
from all subprocesses $q\bar q$ (dot-dashed line), $qg$ (long-dashed line),
$gg$ (dotted line) and $qq$ (short-dashed line). 
\item[Fig. 4a.]
Same as in Fig. 3a but now using GRSV01 valence scenario
and at $p_T=6~{\rm GeV/c}$.
\item[Fig. 4b.]
Same as in Fig. 3b but now using GRSV01 valence scenario
and at $p_T=6~{\rm GeV/c}$.
\item[Fig. 5.]
The dependence of $d^2\Delta\sigma/dQ~dp_T$ on $p_T$
at $\sqrt{s}=200$ GeV with the GRSV01 standard scenario and $\mu^2=Q^2+p_T^2$.
Reading from top to bottom the solid lines are the NLO results
for $Q=2$, 6, and 10 ${\rm GeV/c^2}$ and the dashed lines are the LO results. 
\item[Fig. 6.]
The dependence of $d^3\Delta\sigma/dQ~dp_T~dy$ on $y$ at $\sqrt{s}=200$ GeV 
and $p_T=2~{\rm GeV/c}$ with the GRSV01 standard scenario and
$\mu^2=Q^2+p_T^2$.
Reading from top to bottom the solid lines are the NLO results
for $Q=2$, 6, and 10 ${\rm GeV/c^2}$ and the dashed lines are the LO results. 
\item[Fig. 7a.]
The quantity $N(p_T,\mu/\mu_0)$ (Eq. (\ref{eqn4.8})) plotted in the range
$0.2<\mu/\mu_0<5$ (logarithmic scale) with $Q=8~{\rm GeV/c^2}$ 
and $\mu_0^2=Q^2+p_T^2$. The results are shown for
$p_T=2~{\rm GeV/c}$ (solid line), $p_T=10~{\rm GeV/c}$ (dashed line),
$p_T=20~{\rm GeV/c}$ (dot-dashed line).
The three upper curves on the left hand side are the LO results whereas
the three lower curves are the NLO results.
\item[Fig. 7b.]
The quantity $N(p_T,y,\mu/\mu_0)$ (analogous to Eq. (\ref{eqn4.8})) plotted 
at $p_T=2~{\rm GeV/c}$ in the range $0.2<\mu/\mu_0<5$ (logarithmic scale)
with $Q=8~{\rm GeV/c^2}$ and $\mu_0^2=Q^2+p_T^2$. The results are shown for
$y=\pm 1$ (solid line) and $y=0$ (dashed line).
The two upper curves on the left hand side are the LO results whereas
the two lower curves are the NLO results.
\item[Fig. 8a.]
The $K$-factor (Eq. (\ref{eqn4.9})) for $d^2\Delta\sigma/dQ~dp_T$ plotted 
in the range $2<p_T<30~{\rm GeV/c}$ for $\mu^2=Q^2+p_T^2$. The values for $Q$
are $Q=2~{\rm GeV/c^2}$ (solid line),  
$Q=6~{\rm GeV/c^2}$ (dashed line) and
$Q=10~{\rm GeV/c^2}$ (dot-dashed line). 
\item[Fig. 8b.]
The $K$-factor (Eq. (\ref{eqn4.9})) for $d^3\Delta\sigma/dQ~dp_T~dy$ 
at $p_T=2~{\rm GeV/c}$ with $\mu^2=Q^2+p_T^2$. The values for $Q$ are
$Q=2~{\rm GeV/c^2}$ (solid line),
$Q=6~{\rm GeV/c^2}$ (dashed line) and
$Q=10~{\rm GeV/c^2}$ (dot-dashed line).
\item[Fig. 9a.]
The LO longitudinal asymmetry $A_{\rm LL}(p_T)$ (Eq. (\ref{eqn4.7}))
in $\%$ plotted in the range
$2<p_T<30~{\rm GeV/c}$ for $Q=6~{\rm GeV/c^2}$ and $\mu^2=Q^2+p_T^2$.
GRSV01, standard scenario (solid line),
GRSV01, valence scenario (dashed line),
BB, set 1 (dot-dashed line) and BB, set 2 (dotted line).
\item[Fig. 9b.]
Same as in Fig. 9a but now for $A_{\rm LL}(p_T)$ (Eq. (\ref{eqn4.7})) in NLO.
\item[Fig. 9c.]
The ratio of $A_{\rm LL}(p_T)$ in NLO in Fig. 9b divided by 
$A_{\rm LL}(p_T)$ in LO in Fig. 9a.
\item[Fig. 10a.]
The LO longitudinal asymmetry $A_{\rm LL}(p_T,y)$ (Eq. (\ref{eqn4.7})) in $\%$ 
at $p_T=6~{\rm GeV/c}$ for $Q=6~{\rm GeV/c^2}$ and $\mu^2=Q^2+p_{T}^2$.
GRSV01, standard scenario (solid line),
BB, set 1 (dot-dashed line) and BB, set 2 (dotted line).
\item[Fig. 10b.]
Same as in Fig. 10a but now for $A_{\rm LL}(p_T,y)$ (Eq. (\ref{eqn4.7})) in NLO.
\item[Fig. 10c.]
The ratio of $A_{\rm LL}(p_T,y)$ in NLO in Fig. 10b divided by 
$A_{\rm LL}(p_T,y)$ in LO in Fig. 10a.
\end{description}

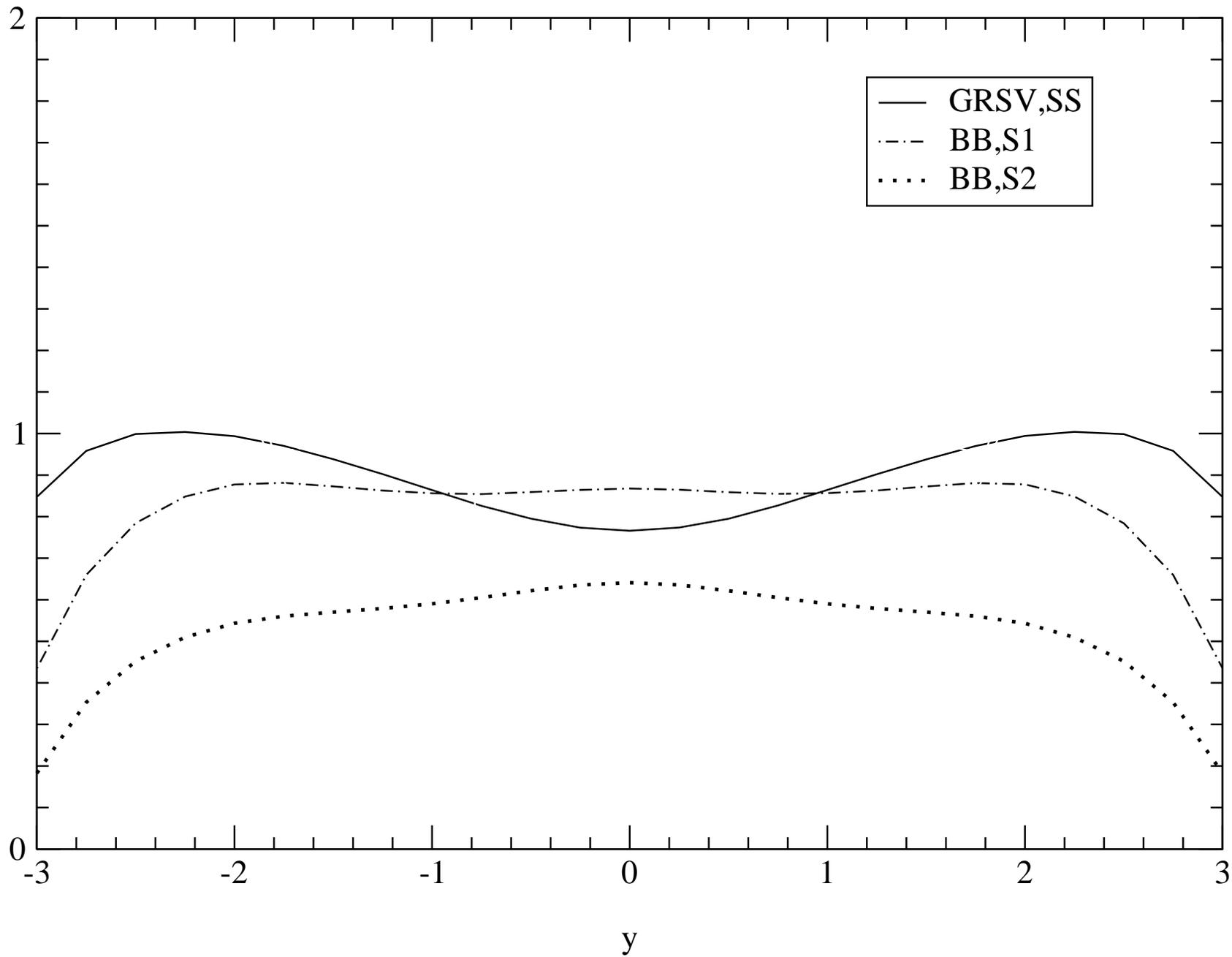
\end{document}